\documentclass[aps,prb,twocolumn,amsmath,amssymb,showpacs,superscriptaddress]{revtex4-1}

\usepackage{graphicx}
\usepackage{amsmath,amssymb,amsfonts}
\usepackage{textcomp}
\usepackage{hyperref}
\usepackage{gensymb}
\usepackage{verbatim}
\usepackage{amsmath}
\hypersetup{breaklinks=true,colorlinks=true,urlcolor=black}
\usepackage{color}
\usepackage{soul}
\DeclareGraphicsExtensions{.jpg,.pdf,.png,.eps}

\begin{document}
\title{Anisotropic thermodynamic and transport properties of single crystalline CaKFe$_{4}$As$_{4}$}
\author{W. R. Meier}
\affiliation{Ames Laboratory US DOE, Iowa State University, Ames, Iowa 50011, USA}
\affiliation{Department of Physics and Astronomy, Iowa State University, Ames, Iowa 50011, USA}
\author{T. Kong}
\affiliation{Ames Laboratory US DOE, Iowa State University, Ames, Iowa 50011, USA}
\affiliation{Department of Physics and Astronomy, Iowa State University, Ames, Iowa 50011, USA}
\author{U. S. Kaluarachchi}
\affiliation{Ames Laboratory US DOE, Iowa State University, Ames, Iowa 50011, USA}
\affiliation{Department of Physics and Astronomy, Iowa State University, Ames, Iowa 50011, USA}
\author{V. Taufour}
\affiliation{Ames Laboratory US DOE, Iowa State University, Ames, Iowa 50011, USA}
\author{N. H. Jo}
\affiliation{Ames Laboratory US DOE, Iowa State University, Ames, Iowa 50011, USA}
\affiliation{Department of Physics and Astronomy, Iowa State University, Ames, Iowa 50011, USA}
\author{G. Drachuck}
\affiliation{Ames Laboratory US DOE, Iowa State University, Ames, Iowa 50011, USA}
\affiliation{Department of Physics and Astronomy, Iowa State University, Ames, Iowa 50011, USA}
\author{A. E. B\"{o}hmer}
\affiliation{Ames Laboratory US DOE, Iowa State University, Ames, Iowa 50011, USA}
\author{S. M. Saunders}
\affiliation{Ames Laboratory US DOE, Iowa State University, Ames, Iowa 50011, USA}
\affiliation{Department of Physics and Astronomy, Iowa State University, Ames, Iowa 50011, USA}
\author{A. Sapkota}
\affiliation{Ames Laboratory US DOE, Iowa State University, Ames, Iowa 50011, USA}
\affiliation{Department of Physics and Astronomy, Iowa State University, Ames, Iowa 50011, USA}
\author{A. Kreyssig}
\affiliation{Ames Laboratory US DOE, Iowa State University, Ames, Iowa 50011, USA}
\affiliation{Department of Physics and Astronomy, Iowa State University, Ames, Iowa 50011, USA}
\author{M. A. Tanatar}
\affiliation{Ames Laboratory US DOE, Iowa State University, Ames, Iowa 50011, USA}
\affiliation{Department of Physics and Astronomy, Iowa State University, Ames, Iowa 50011, USA}
\author{R. Prozorov}
\affiliation{Ames Laboratory US DOE, Iowa State University, Ames, Iowa 50011, USA}
\affiliation{Department of Physics and Astronomy, Iowa State University, Ames, Iowa 50011, USA}
\author{A. I. Goldman}
\affiliation{Ames Laboratory US DOE, Iowa State University, Ames, Iowa 50011, USA}
\affiliation{Department of Physics and Astronomy, Iowa State University, Ames, Iowa 50011, USA}
\author{Fedor F. Balakirev}
\affiliation{National High Magnetic Field Laboratory, Los Alamos National Laboratory, MS-E536, Los Alamos, New Mexico 87545, USA}
\author{Alex Gurevich}
\affiliation{Department of Physics, Old Dominion University, Norfolk, Virginia 23529, USA}
\author{S. L. Bud'ko}
\affiliation{Ames Laboratory US DOE, Iowa State University, Ames, Iowa 50011, USA}
\affiliation{Department of Physics and Astronomy, Iowa State University, Ames, Iowa 50011, USA}
\author{P. C. Canfield}
\affiliation{Ames Laboratory US DOE, Iowa State University, Ames, Iowa 50011, USA}
\affiliation{Department of Physics and Astronomy, Iowa State University, Ames, Iowa 50011, USA}
\email[]{canfield@ameslab.gov}

\date{\today}

\begin{abstract}
Single crystalline, single phase CaKFe$_{4}$As$_{4}$ has been grown out of a high temperature, quaternary melt. Temperature dependent measurements of x-ray diffraction, anisotropic electrical resistivity, elastoresistivity, thermoelectric power, Hall effect, magnetization and specific heat, combined with field dependent measurements of electrical resistivity and field and pressure dependent measurements of magnetization indicate that CaKFe$_{4}$As$_{4}$ is an ordered, stoichiometric, Fe-based superconductor with a superconducting critical temperature, $T_c$~=~35.0\,$\pm$\,0.2\,K. Other than superconductivity, there is no indication of any other phase transition for 1.8\,K~$\leq$~$T$~$\leq$~300\,K. All of these thermodynamic and transport data reveal striking similarities to that found for optimally- or slightly over-doped (Ba$_{1-x}$K$_x$)Fe$_2$As$_2$, suggesting that stoichiometric CaKFe$_4$As$_4$ is intrinsically close to what is referred to as "optimal-doped" on a generalized, Fe-based superconductor, phase diagram. The anisotropic superconducting upper critical field, $H_{c\text{2}}(T)$, of CaKFe$_{4}$As$_{4}$ was determined up to 630 kOe. The anisotropy parameter $\gamma(T)=H_{c\text{2}}^{\perp}/H_{c\text{2}}^{\|}$, for $H$ applied perpendicular and parallel to the c-axis, decreases from $\simeq 2.5$ at $T_c$ to $\simeq 1.5$ at 25 K which can be explained by interplay of paramagnetic pairbreaking and orbital effects. The slopes of $dH_{c\text{2}}^{\|}/dT\simeq-44$ kOe/K and $dH_{c\text{2}}^{\perp}/dT \simeq-109$ kOe/K at $T_c$ yield an electron mass anisotropy of $m_{\perp}/m_{\|}\simeq 1/6$ and short Ginzburg-Landau coherence lengths $\xi_{\|}(0)\simeq 5.8\,\text{\AA}$ and $\xi_{\perp}(0)\simeq 14.3\,\text{\AA}$. The value of $H_{c\text{2}}^{\perp}(0)$ can be extrapolated to $\simeq 920$ kOe, well above the BCS paramagnetic limit.

\end{abstract}

\maketitle 

\section{INTRODUCTION}
BaFe$_{2}$As$_{2}$ has become one of the archetypical examples of Fe-based superconductivity\cite{Canfield10}. It was the first of the \textit{Ae}Fe$_{2}$As$_{2}$ (122) structures (\textit{Ae} = Ba, Sr, Ca) found to support superconductivity (with K-substitution for Ba) \cite{Rotter08b} and was almost immediately studied in single crystalline form \cite{Ni08a}. With the discovery that cobalt substitution on the iron site could stabilize superconductivity \cite{Sefat08}, extensive studies of Ba(Fe$_{1-x}$\textit{TM}$_{x}$)$_{2}$As$_{2}$ (\textit{TM} = transition metal) series revealed the basic relations between the structural, magnetic and electronic degrees of freedom in these compounds as substitutions progressed from under-doped (rising $T_c$, often coexisting with antiferromagnetic (AFM) order), to optimally-doped (maximal $T_c$ near the disappearance of the AFM transition), to over-doped (dropping $T_c$ in paramagnetic state).\cite{Canfield10} SrFe$_{2}$As$_{2}$ was also studied, albeit to a lesser degree for selected substitutions on the \textit{Ae} as well as \textit{TM} sites. \cite{Johnston10,Stewart11} 

CaFe$_{2}$As$_{2}$, on the other hand, has proven more difficult to modify with substitution. CaFe$_{2}$As$_{2}$ is the lightest member of the \textit{Ae}Fe$_{2}$As$_{2}$ series and was discovered in single crystalline form \cite{Ni08b} only after the discovery of Fe-based superconductivity. \cite{Rotter08b,Kamihara08} CaFe$_2$As$_2$ manifests a strongly coupled, first order, magnetic and structural phase transition \cite{Canfield09} and pressure studies of CaFe$_{2}$As$_{2}$ led to the discovery of a collapsed tetragonal (cT) phase for $p$~$\textgreater$~0.4\,GPa \cite{Canfield09,Kreyssig08} and a much wider appreciation of the proximity and influence of the cT phase in all of the \textit{Ae}Fe$_{2}$As$_{2}$ compounds. Systematic studies of transition metal substitution in the Ca(Fe$_{1-x}$\textit{TM}$_{x}$)$_{2}$As$_{2}$ series were only possible once the extreme strain and pressure sensitivity of the CaFe$_{2}$As$_{2}$ host were appreciated and it was realized that internal strain had to be controlled by careful post-growth annealing of single crystalline samples\cite{Ran11,Ran12,Ran14a,Ran14b}. The incredible richness and sensitivity of the CaFe$_{2}$As$_{2}$ system is attributed to the Ca-ions being at or near the edge of the steric tolerance of the un-collapsed 122 structure.

Recently, another consequence of the size of the Ca-ions has been discovered. Iyo et al.\cite{Iyo16} have found that a family of ordered Ca\textit{A}Fe$_{4}$As$_{4}$ (1144) compounds can be formed for \textit{A} = K, Rb, Cs where the key to the formation is the difference in ionic size between the Ca and the \textit{A} ion. This family is not a (Ca$_{1-x}$\textit{A}$_{x}$)Fe$_{2}$As$_{2}$ solid-solution, where the Ca and \textit{A} ions randomly occupy a single crystallographic site, \cite{Wang13} but rather is a distinct, quaternary, line compound in which the Ca and \textit{A} sites form alternating planes along the crystallographic \textit{c}-axis, separated by FeAs slabs\cite{Iyo16}. In essence, the Ca\textit{A}Fe$_{4}$As$_{4}$ structure is identical to the CaFe$_{2}$As$_{2}$ structure, just with layer by layer segregation of the Ca and \textit{A} ions. The 1144 structure was also found for Sr\textit{A}Fe$_{4}$As$_{4}$ (\textit{A} = Rb, Cs). Solid-solutions of Ca (Sr) 122 structures were found for \textit{A} = Na (Na, K) respectively as well as for all attempted Ba-based systems. \cite{Iyo16} Superconducting transition temperatures ($T_c$) were inferred from both resistance and magnetization data with $T_c$ values ranging between 31 and 37\,K. \cite{Iyo16} These $T_c$-values are among the highest reported for bulk, fully ordered, stoichiometric Fe-based superconductors. As such, CaKFe$_{4}$As$_{4}$ offers a unique opportunity to study relatively high transition temperature, Fe-based superconductivity in a highly ordered compound, at ambient pressure.

In their discovery paper, Iyo et al. synthesized and studied polycrystalline samples. \cite{Iyo16} A vital next step is to grow and study single crystalline samples so the details of the intrinsic properties, including anisotropies, can be examined. In this paper we outline experimental details for the growth of single phase, single crystalline CaKFe$_{4}$As$_{4}$ and present structural, thermodynamic, and transport data as a function of temperature, field, and pressure. We find that CaKFe$_{4}$As$_{4}$ is a rare example of an ordered Fe-based superconductor that appears to be intrinsically near optimally- or slightly over-doped and has a $T_c$ value of 35.0\,$\pm$\,0.2\,K.

\section{CRYSTAL GROWTH AND EXPERIMENTAL METHODS}
CaKFe$_{4}$As$_{4}$ single crystals were grown by high temperature solution growth out of FeAs flux in a manner similar to CaFe$_{2}$As$_{2}$ and K$_{2}$Cr$_{3}$As$_{3}$. \cite{Ran11,Kong15a} Lump, elemental K (Alfa Aesar 99.95\%) and distilled Ca pieces (Ames Laboratory, Material Preparation Center (MPC) $\textgreater$\,99.9\%) were combined with ground, pre-reacted Fe$_{0.512}$As$_{0.488}$ precursor in a ratio of K~:~Ca~:~Fe$_{0.512}$As$_{0.488}$ = 1.2~:~0.8~:~20, with a total mass of roughly two grams, in a fritted, alumina crucible set (Canfield Crucible Set, or CCS). \cite{Canfield16} The precursor was synthesized from As (Alfa Aesar 99.9999\%) and Fe (Alfa Aesar 99.9+\%) powders in a 1~:~1.05 atomic ratio in an argon filled fused-silica ampoule. \cite{Ran14b} The filled CCS was welded into a Ta crucible which itself was sealed into a fused-silica ampoule\cite{Kong15a}. The growth ampoule was heated over 1 hour to 650\textdegree C, held for 3 hours then heated over 2 hours to 1180\textdegree C, held at this temperature for 5 hours, cooled to 1050\textdegree C over 2 hours, and then slowly cooled from 1050\textdegree C to 930\textdegree C over 30 hours. When this final temperature was reached, the assembly was removed from the furnace, inverted into a centrifuge and spun to expedite the separation of crystals from the liquid flux\cite{Canfield92,Canfield10b}.

Single crystalline CaKFe$_{4}$As$_{4}$ grows as mirror-like, metallic, micaceous plates of 0.1-0.2\,mm thickness which can, in some cases, be limited in area by the inner diameter of the crucible (see inset to Fig.\,\ref{anisoRT}, below). The crystallographic \textit{c}-axis is perpendicular to the plate surface. Single crystals of CaKFe$_{4}$As$_{4}$ are not particularly air sensitive and can remain in air for weeks without any noticeable degradation of their appearance or physical properties.

CaFe$_{2}$As$_{2}$ and KFe$_{2}$As$_{2}$ can be second phases in such growths and care had to be taken in optimizing our final growth protocol as well as in selecting our crystals to be sure that we have little or no amount of either of these phases. The $\sim$170\,K phase transition of CaFe$_{2}$As$_{2}$ \cite{Canfield09} is most apparent in temperature dependent resistance measurements and the low temperature superconducting phase transition in KFe$_{2}$As$_{2}$ ($T_{c}$ = 3.8 K\cite{Rotter08a}), as seen in the low field magnetization measurement, is the most sensitive way to detect its presence. All samples used in these studies were screened for both impurity phases. A more detailed discussion of how crystal growth was optimized to the current protocol, in part by minimizing these diagnostic signatures of second phases, will be presented in a separate paper.

Single crystals of CaKFe$_{4}$As$_{4}$ are soft, malleable, and not amenable to grinding for powder x-ray diffraction measurements. In this sense, CaKFe$_{4}$As$_{4}$ is mechanically more similar to CaFe$_2$As$_2$\cite{Ni08b} than to BaFe$_2$As$_2$\cite{Ni08a}. Diffraction measurements on a single crystal were carried out in-house using a Rigaku MiniFlex II powder diffractometer in a Bragg-Brentano geometry\cite{Jesche16} with a Cu K$_{\alpha}$ source and a graphite monochromator in front of the detector. Single crystal high-energy x-ray diffraction measurements were made at station 6-ID-D at the Advanced Photon Source (APS) using an x-ray wavelength of $ \lambda $~=~0.123589\,$\text{\normalfont\AA}$ and a beam size of 100\,$\times$\,100\,$\mu$ m$ ^{2}$. The synchrotron measurements were performed on a 0.5\,$\times$\,0.5\,$\times$\,0.05 mm$^{3}$ sample using a He, closed-cycle, refrigerator. Three Be domes were placed over the sample and evacuated with the middle one functioning as heat shield, and a small amount of He gas was added to the inner dome to facilitate thermal coupling. The cryostat was mounted to the sample stage of a 6-circle diffractometer, and a MAR345 image plate was positioned 1.487\,m behind the sample to measure the diffracted x-rays transmitted through the sample spanning a scattering angle of $|$2$\theta |$ $\leq$ 6.65\textdegree . Data were taken by recording an image while tilting the sample along two rocking angles perpendicular to the incident x-ray beam \cite{Kreyssig07}. $(hk0)$ and $(h0\ell)$ reciprocal planes were recorded for temperatures from 300\,K down to 6\,K.

Temperature and field dependent magnetization, resistance and specific heat measurements were carried out using Quantum Design (QD), Magnetic Property Measurement Systems (MPMS) and Physical Property Measurement Systems (PPMS). Temperature dependent specific heat measurements taken for $H\|c$ in applied magnetic field resulted in significant torque on the thin, plate-like samples. Even with care, the calorimeter platform rotated by $\lesssim$\,10\textdegree~as a result of measurements in applied field up to 140\,kOe, and in some cases there was a loss of some sample mass due to exfoliation. As a result, specific heat data measured in applied fields are shown normalized to the zero-field data in the normal state. Hall resistivity data were collected using the AC transport option of a QD PPMS in a four-wire geometry with switching the polarity of the magnetic field $H\|c$ to remove any magnetoresistive components due to misalignment of the voltage contacts. Thermoelectric power (TEP) measurements were performed using a DC, alternating temperature gradient technique \cite{Mun10b} with the temperature-field environment provided by a QD PPMS unit.

Optical imaging of the magnetic flux distribution was performed in a $^4$He flow - type cryostat by using the magneto-optical Faraday effect. In the experiment a transparent bismuth-doped iron-garnet ferrimagnetic ``indicator" film with in-plane magnetization is placed directly on top of the sample. In the images, brightness is proportional to the value of $B_z (\vec{r})$ with black level set at $B_z (\vec{r})=0$ and the colors are related to the absolute orientation of $B_z (\vec{r})$: green for out of page and yellow for into the page directions for our setup. More details on the technique and magneto-optical imaging of other Fe-based superconductors can be found elsewhere \cite{Prozorov2009d,Prozorov2008a,Prozorov2010a}.

The pressure dependence of $T_c$ was determined by measurements of pressure dependent magnetization. Data up to 1.2\,GPa were taken in a commercial, piston-cylinder, HMD cell using Daphne 7373 as pressure medium and Pb as a manometer \cite{Schilling81}.  Data for $p$~$\textless$~4.0\,GPa were taken using a moissanite anvil cell\cite{Alireza07} using Daphne 7474 as pressure medium and utilizing ruby fluorescence at 77\,K as a manometer. For both pressure cells, the temperature-field environment was provided by a QD MPMS unit.

The samples for anisotropic resistivity measurements were cleaved from larger crystals with sides along $\langle100\rangle$ directions using a razor blade. Samples used for inter-plane ($I\,\|\,c$) measurements typically had dimensions of 0.5\,$\times$\,0.5\,$\times$\,0.02\,mm$^3$ ($a\times b \times c$). The samples for in-plane ($I\perp c$) measurements were typically of 1.5\,$\times$\,0.2\,$\times$\,0.02\,mm$^3$ size. Contacts for standard four probe, in-plane resistivity measurements were soldered using Sn \cite{Tanatar09,Tanatar10,Tanatar}. For inter-plane resistivity measurements we used two-probe measurements, relying on the negligible contact resistance. The top and bottom surfaces of the samples were covered with Sn solder \cite{Tanatar10,Tanatar} and 50\,$\mu$m silver wires were attached to enable measurements in a four-probe configuration.  Soldering produced contacts with typical resistances in the 10\,$\mu\Omega$ range. Inter-plane resistivity was measured using a two-probe technique with currents in 1 to 10\,mA range (depending on sample resistance which is typically 1\,m$\Omega$).  A four-probe scheme was used to measure the sample resistance, $R_s$, and contact resistance, $R_c$, in series. Taking into account that $R_s\,\gg\,R_c$, contact resistance represents a minor correction of the order of 1 to 5\%. This can be directly seen for our samples for temperatures below the superconducting $T_c$, where $R_s\,=\,0$ and the measured resistance represents $R_c$ \cite{Tanatar09,Tanatar10}. The details of the measurement procedure can be found in Ref.~\onlinecite{Tanatar09b}. 

$H_{c\text{2}}(T)$ was determined via magnetoresistance measurements with $I\perp c$. Both DuPont 4929N silver paint and Epotek-H20E silver epoxy were used to attach contact leads onto the samples (Pt for measurements static field measurements and twisted copper wires for pulsed field measurements). For static fields below 140 kOe, resistance was measured using a QD PPMS-14 ($T$ = 1.8-305 K, $H$ = 0-140 kOe, $f$ = 17 Hz.). Higher field data were obtained in a 630 kOe  pulsed magnet at the National High Magnetic Field Laboratory (NHMFL), Los Alamos, using a high-frequency, synchronous digital lock-in technique ($f$ = 148 kHz).

Elastoresistivity was measured using a piezostack-based setup, similar to that described in Refs.~\onlinecite{Chu12, Kuo13}. Samples of approximate dimensions, 1\,$\times$\,0.3\,$\times$\,0.04\,mm$^3$, were glued on one side of a “Piezomechanik GmbH PSt 150/5x5/7” piezostack, as shown in the inset in Fig.~\ref{elastoresistivity} below. The change of sample resistance was measured with “Lakeshore Model 372” AC Resistance Bridge as a function of anisotropic strain, monitored in situ using crossed strain gauges glued to the opposite side of the piezostack. The temperature environment was provided by a Janis SHI-950-T closed cycle cryostat.

\begin{figure}
	\includegraphics[scale=1]{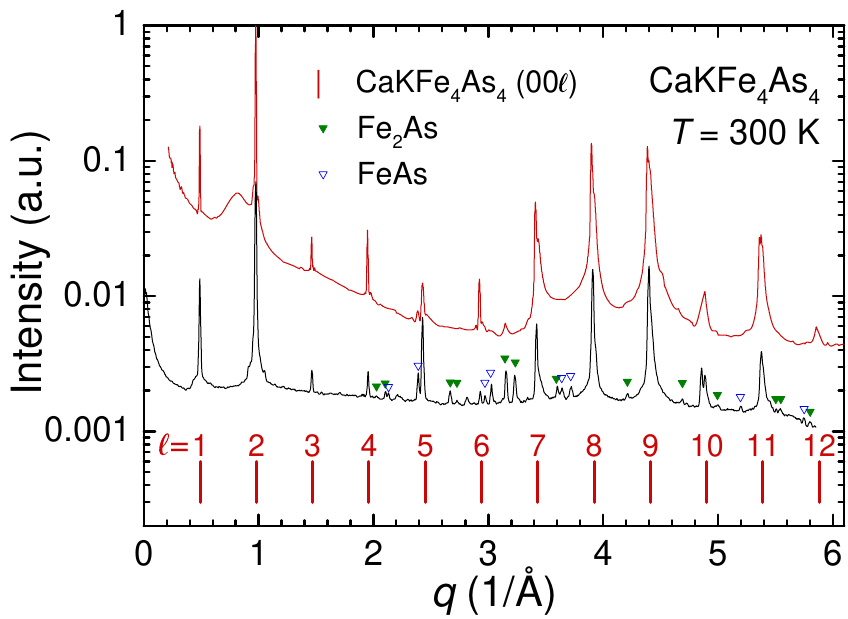}%
	\caption{(Color online) x-ray diffraction data showing (00$\ell$) diffraction peaks from in-lab diffraction measurements on a single crystalline plate (upper data set) and high-energy x-ray diffraction measurements taken at APS (lower data set). Note that $\ell$ = odd (00$\ell$) lines are consistent with the ordered CaKFe$_{4}$As$_{4}$ structure and are formally forbidden in a (Ca$_{1-x}$K$_{x}$)Fe$_{2}$As$_{2}$ structure \cite{Iyo16}.
		\label{xrd}}
\end{figure}

\section{EXPERIMENTAL RESULTS}
\begin{figure}
	\includegraphics[scale=1]{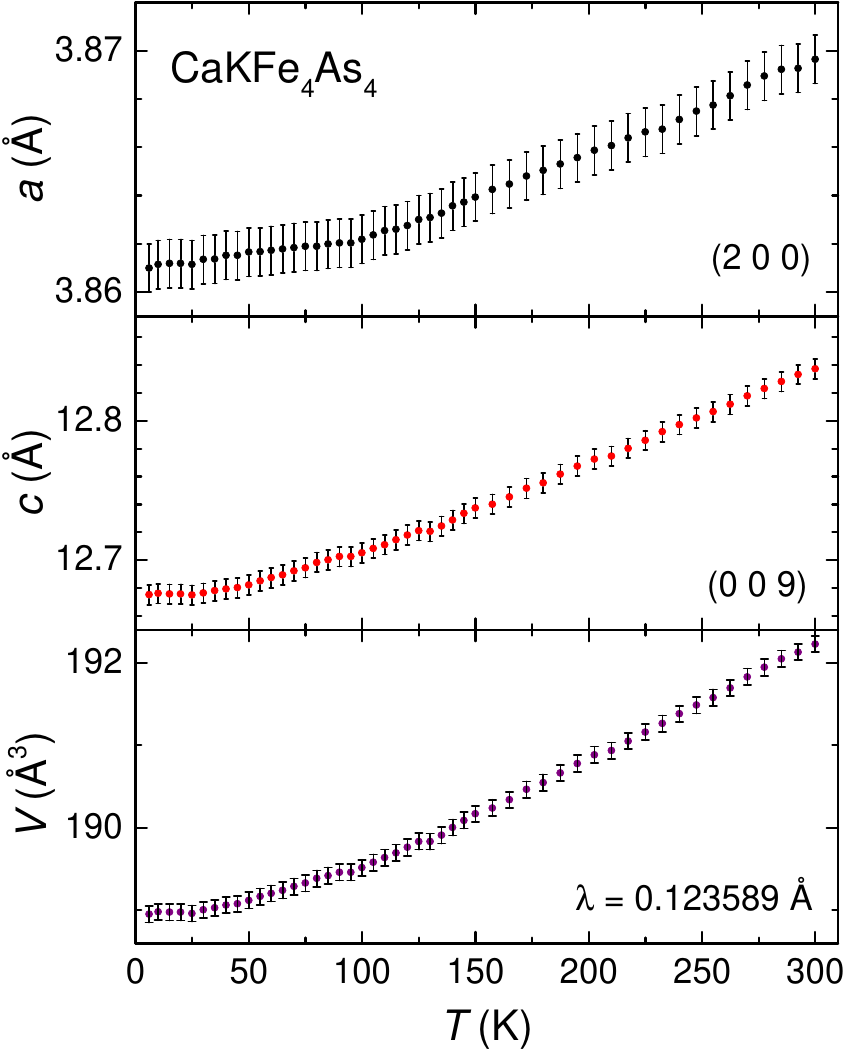}%
	\caption{(Color online) Temperature dependence of CaKFe$_{4}$As$_{4}$ \textit{a}- and \textit{c}-lattice parameters as well as unit cell volume as determined from (200) and (009) diffraction lines measured via high-energy x-ray diffraction.
		\label{latticeT}}
\end{figure}

Figures~\ref{xrd} and~\ref{latticeT} present x-ray diffraction data and the temperature dependence of the CaKFe$_{4}$As$_{4}$ unit cell dimensions and volume, respectively. The presence of $h+k+\ell$ = odd peaks, which are forbidden for the $I4/mmm$, \textit{Ae}Fe$_{2}$As$_2$ structure, indicates that, instead, CaKFe$_{4}$As$_{4}$ assumes the ordered $P4/mmm$ structure. \cite{Iyo16} Given the relatively large \textit{c}-axis dimension we are able to detect $(00\ell)$ peaks for all $\ell\,\leqslant\,12$ in our in-house unit with Cu K$_{\alpha}$ radiation. The broad peak on the low-$q$ side of the (002) peak in the in-house data set is from a thin film of vacuum grease used to affix the thin CaKFe$_{4}$As$_{4}$ plate to the zero-background single crystalline silicon sample holder. Virtually no signatures of $(00\ell)$ peaks associated with CaFe$_2$As$_2$ or KFe$_2$As$_2$ are found. The agreement between the in-house, Cu K$_{\alpha}$ data, which comes from the surface of the crystalline plate, and the high-energy x-ray data, which penetrates through the bulk of the sample, indicates that the sample is essentially single phase and uniform throughout its whole volume. The other, small, marked peaks are associated with traces (note that data is presented on a log scale) of FeAs and Fe$_{2}$As flux remaining on the sample after decanting. The temperature dependencies of the \textit{a}- and \textit{c}-lattice parameters of the CaKFe$_{4}$As$_{4}$ sample, measured using high energy x-rays at the APS, are both monotonic and decrease with decreasing temperature. There is no evidence of a structural phase transition over our measured 6\,K\,$ \textless\,T\,\textless$\,300\,K temperature range. The room temperature lattice parameters are close to reported values for polycrystalline samples\cite{Iyo16} (a = 3.866 $\text{\AA}$, c = 12.817 $\text{\AA}$) as well as for single crystal samples\cite{Mou16} (a = 3.8659 $\text{\AA}$, c = 12.884 $\text{\AA}$)

\begin{figure}
	\includegraphics[scale=1]{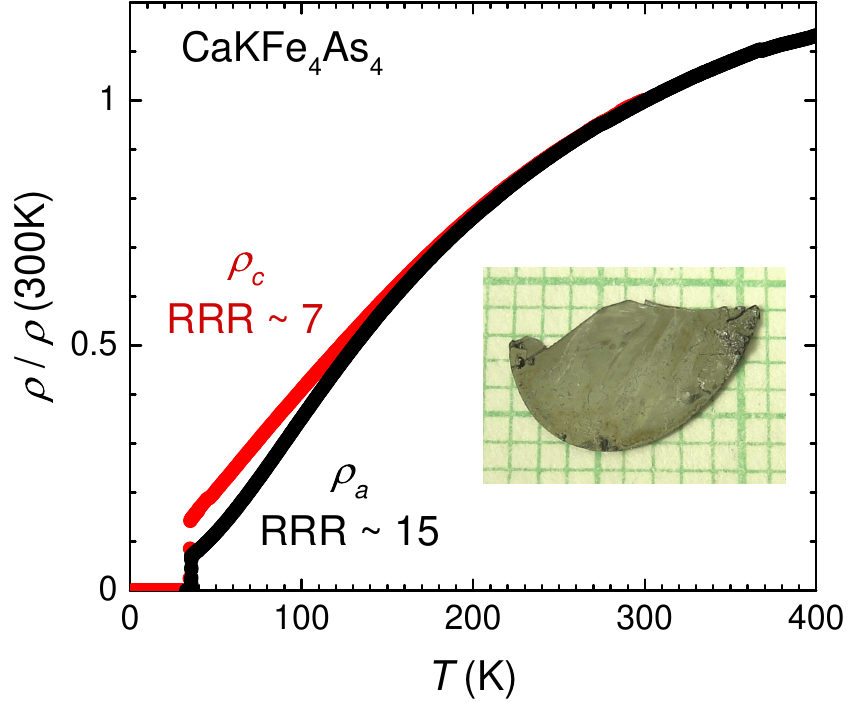}%
	\caption{(Color online) Temperature-dependent in-plane, $\rho_a(T)$, and inter-plane, $\rho_c(T)$, resistivity of CaKFe$_4$As$_4$, plotted using normalized resistivity scales, $\rho(T)/\rho(300\,K)$. At 300\,K, $\rho_a$ $\sim$ 300\,$\mu\Omega$\,cm and $\rho_c$ $\sim$ 1000-2000\, $\mu\Omega$\,cm. Inset: picture of a CaKFe$_4$As$_4$ single crystal shown over a mm-grid.
		\label{anisoRT}}
\end{figure}

\begin{figure}
	\includegraphics[scale=1]{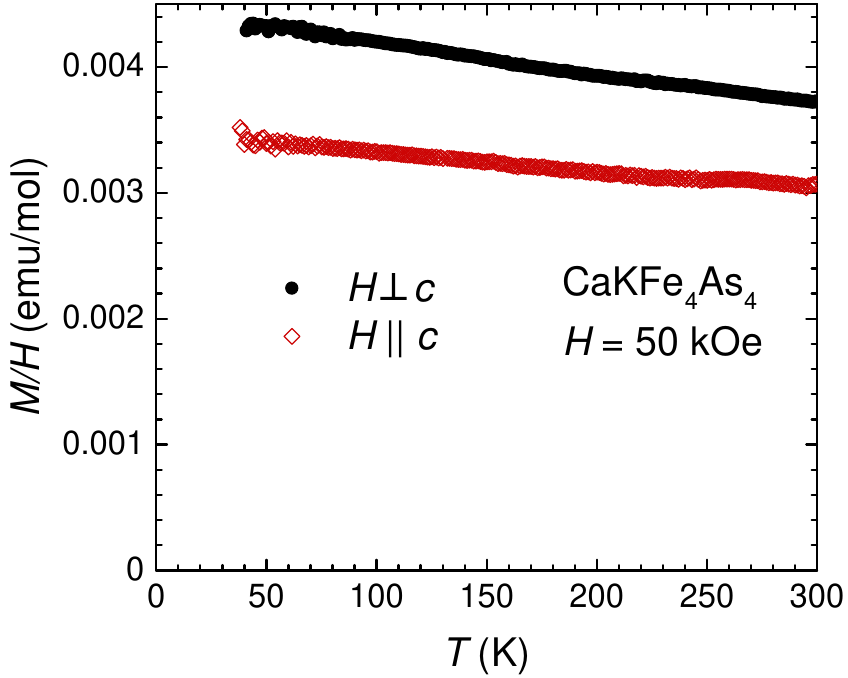}%
	\caption{(Color online) Anisotropic, temperature dependent magnetization divided by applied field ($M(T)/H$) of CaKFe$_{4}$As$_{4}$ taken for $H$~=~50\,kOe applied along the crystallographic \textit{c}-axis and perpendicular to the crystallographic \textit{c}-axis. Due to $T_c$ at 35\,K, data shown are for 40\,K~$<$~$T$~$<$~300\,K.  
		\label{MTaboveTc}}
\end{figure}

The anisotropic, temperature dependent, normalized electrical resistivity and magnetization of CaKFe$_{4}$As$_{4}$ are shown in Figs.~\ref{anisoRT} and~\ref{MTaboveTc}. In-plane electrical resistance measurements, $\rho_a(T)$, were performed on multiple samples, both with soldered Sn and silver-epoxy contacts, revealing a highly reproducible temperature dependence. We also performed measurements of $\rho_c(T)$ on two samples and obtained qualitatively similar temperature dependencies of the electrical resistivity. CaKFe$_{4}$As$_{4}$ manifests very similar temperature dependencies of $\rho_a$ and $\rho_c$ with only slight differences for $T$ $\textless$ 150 K. We find residual resistivity ratios (RRR\,=\,$\rho$(300\,K)/$\rho$(35\,K)) of 15 and 7 for $\rho_a$ and $\rho_c$ respectively. Although we present the electrical resistivity data as normalized, for ease of comparison, we could estimate the room temperature resistivities of $\rho_a$ $\sim$ 300\,$\mu\Omega$\,cm and $\rho_c$ $\sim$ 1000-2000\,$\mu\Omega$\,cm. These values imply that the resistivity value measured on polycrystalline samples ($\rho$(300 K) $\sim$ 3500 $\mu\Omega$\,cm\cite{Iyo16}) may suffer from grain boundary, or other, scattering. The anisotropic $M(T)/H$ data was collected at 50\,kOe in order to allow for adequate signal from a thin, single crystalline plate. The $H\perp c$ data are roughly 15\% larger than the $H\|c$ data and both data sets manifest a weak, essentially linear increase upon cooling from 300\,K to just above $T_c$. For 35\,K~$\textless~T~\textless$~300\,K, neither the temperature dependent electrical resistivity nor the magnetization manifest any features that can be associated with a structural, magnetic, or other electronic phase transition.

\begin{figure}
	\includegraphics[scale=1]{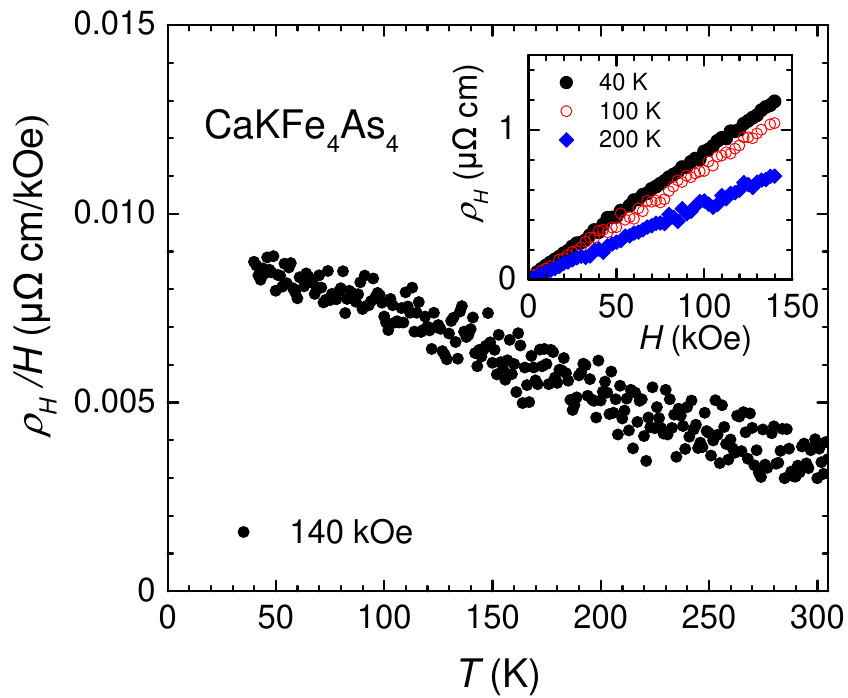}%
	\caption{(Color online) Temperature dependent Hall resistivity divided by field, $\rho_\textrm{H}(T) / H$, (Hall coefficient) of CaKFe$_4$As$_4$ with $H$~=~140\,kOe applied along the crystallographic \textit{c}-axis. Inset shows field dependent Hall resistivity $\rho_\textrm{H}$ at 40, 100, and 200\,K. 
		\label{Hall}}
\end{figure}

\begin{figure}
	\includegraphics[scale=1]{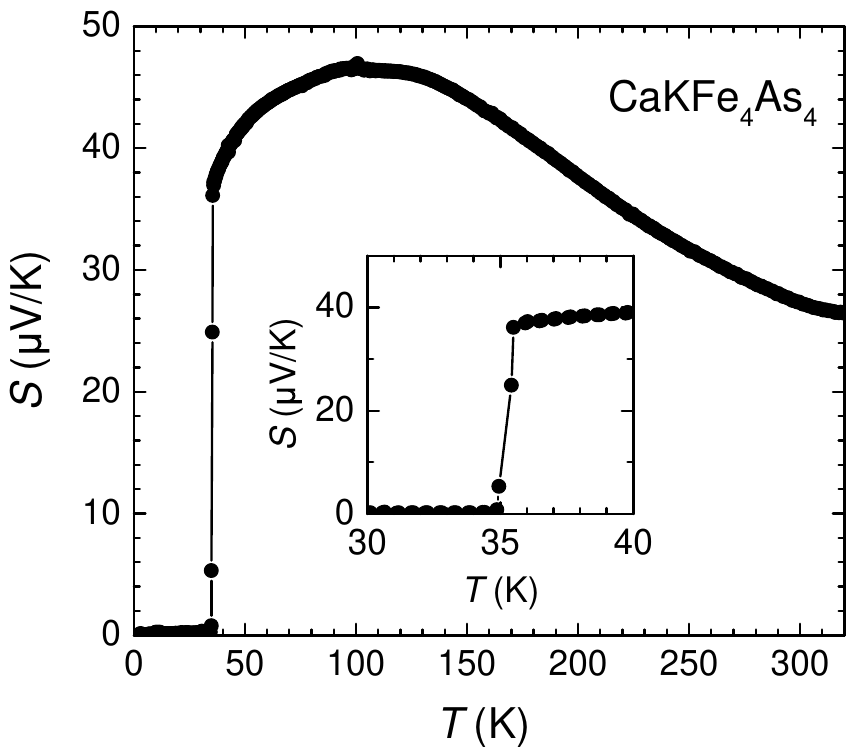}%
	\caption{Temperature dependent thermoelectric power ($S(T)$) for CaKFe$_4$As$_4$ for temperature gradient applied perpendicular to the crystallographic \textit{c}-axis.  
		\label{TEP}}
\end{figure}

Hall resistivity data, as a function of temperature and magnetic field and thermoelectric power data as a function of temperature, $S(T)$, from measurements on CaKFe$_4$As$_4$ are shown in Figs.~\ref{Hall} and~\ref{TEP} respectively. The slope of Hall resistivity (the Hall coefficient) is positive (consistent with the sign of $S(T)$) and linear in field up to the maximum measured field of 140\,kOe. The temperature dependence of the $\rho_H/H$ is weak and close to linear. Although the carrier concentration, roughly evaluated using a single band model, ranges from $\sim$7.4$\times$10$^{21}$\,cm$^{-3}$ at 40\,K to $\sim$1.3$\times$10$^{22}$\,cm$^{-3}$ at 200\,K, CaKFe$_4$As$_4$ undoubtedly has multiple bands\cite{Mou16,Cho16}. Indeed, the temperature-dependent $R_H(T) = \rho_H/H$ shown in Fig. \ref{Hall} is consistent with a multiband electronic structure of CaKFe$_4$As$_4$. In the simplest two-band model, the Hall constant is given by $R_H(T)=(R_1\sigma_1^2+R_1\sigma_2^2)/(\sigma_1+\sigma_2)^2$, where $R_{1,2}$ and $\sigma_{1,2}(T)$ are partial Hall constants and conductivities for band 1 and 2 \cite{Ziman}. Hence, any difference in temperature dependencies of the mean free paths for the electrons/holes in band 1 and 2 would manifest itself in a temperature-dependent $R_H(T)$ even if $R_1=1/q_1n_1$ and $R_2=1/q_2n_2$ are independent of $T$, where $n_1$ and $n_2$ are partial carrier densities in bands 1 and 2, and $q_1$ and $q_2$ are respective charges. 

The thermoelectric power $S(T)$ is near 25\,$\mu$ V/K at room temperature, rises to over 45\,$\mu$ V/K at 100\,K and smoothly drops to near 35 $\mu$ V/K just above $T_c$~=~35\,K, as shown in Fig.~\ref{TEP}. As is the case for the resistivity data, measurements of normal state thermoelectric power for $T \lesssim $ 35 K are precluded by the very large $H_{c2}(T)$ values in the superconducting state (see below). Neither Hall effect nor thermoelectric power data have any features, other than anomaly at $T_c$, that can be associated with any phase transition for 35\,K~$\textless~T~\textless$~300\,K. The overall behaviors of the Hall resistivity and thermoelectric power are similar to those reported for optimally- or slightly over-doped (Ba$_{1-x}$K$_{x}$)Fe$_{2}$As$_{2}$. \cite{Ohgushi12,Halyna14a}.

\begin{figure}
	\includegraphics[scale=1]{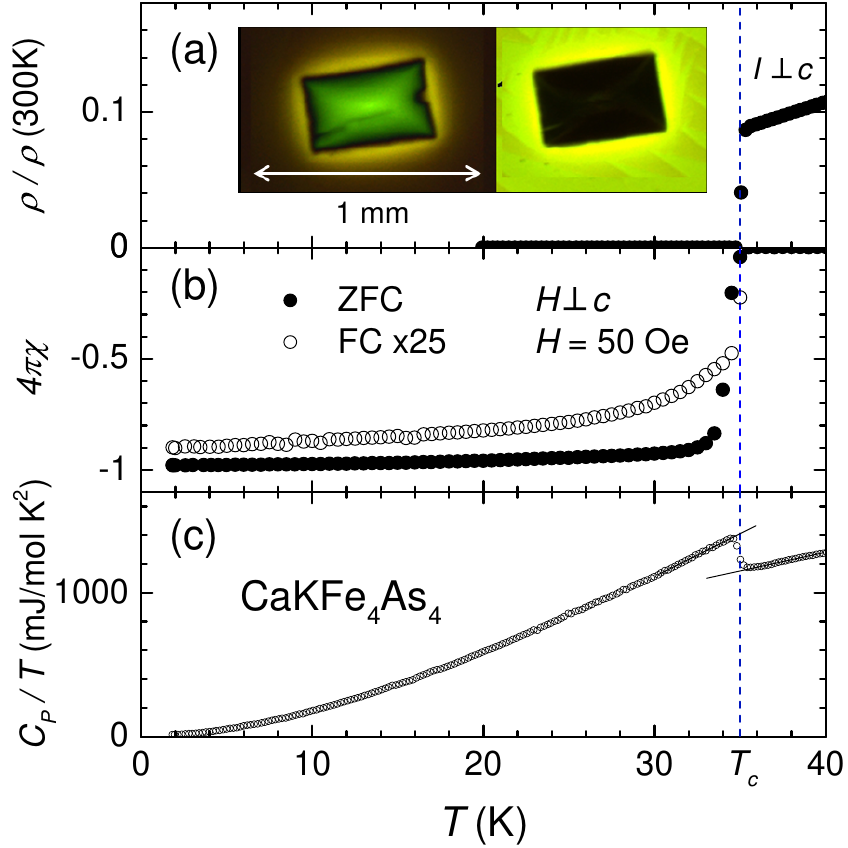}%
	\caption{(Color online) Thermodynamic and transport data taken on CaKFe$_{4}$As$_{4}$ near $T_c$: (a) normalized electrical resistivity. Inset shows the magneto-optic image on a CaKFe$_{4}$As$_{4}$ single crystal (see text for details). (b) FC and ZFC magnetization for $H$~=~50\,Oe for $H$ applied perpendicular to the crystallographic \textit{c}-axis (Note: the FC susceptibility data is multiplied by 25 for clarity), (c) zero field specific heat $ C_p(T)/T $. 
		\label{RChiCpT}}
\end{figure}

Turning now to the superconducting phase transition, Fig.~\ref{RChiCpT} presents the low temperature, normalized, in-plane, electrical resistivity, low field magnetization, and the temperature dependent specific heat. As can be seen, the superconducting phase transition is quite sharp and well defined. In each of these measurements $T_c$\,=\,35.0\,$\pm$\,0.2\,K is the value we can infer from an onset in magnetization, an equi-entropic mid-point in specific heat, and an offset in resistivity. This value is resolvably higher than the $T_c$\,=\,33.1\,K reported by Iyo et al.\cite{Iyo16} We see 1/4$\pi$ shielding in the zero-field-cooled (ZFC) magnetization data; pinning and, as will be discussed below, $\kappa$ are large enough in these samples that we only see a small fraction of a 1/4$\pi$ Meissner effect in the field-cooled (FC) data.  

Magneto-optical imaging of new superconductors is another powerful tool to help confirm the bulk nature of superconductivity via screening of the external magnetic field and study of the vortex physics and irreversible magnetic properties. The inset of Fig.~\ref{RChiCpT} shows magneto-optical imaging of a single crystals of CaKFe$_4$As$_4$. The left image shows magnetic flux trapped by a superconductor due to vortex pinning. In the experiment the sample was cooled in a 1 kOe magnetic field from above $T_c$ to 5 K and then the magnetic field was turned off. Motion of escaping Abrikosov vortices is hindered by pinning centers forming a pyramid - like distribution of vortex density, where height is proportional to $B_z (\vec{r})$. This is so-called remanent ``Bean" critical state\cite{Bean64}. The right image shows state of the sample after it was cooled without applied field from above $T_c$ to 5 K at which point a 220 Oe magnetic field was applied. This is superconducting shielding that mostly probes Meissner screening, which at this low field is about 100\%. (A $H_{c1}$ value of approximately 440 Oe was obtained from London penetration depth measurements.) \cite{Cho16}.

Our magneto-optical and magnetization data show that CaKFe$_4$As$_4$ exhibits a classical irreversible magnetic behavior close to the critical state of a strong type-II superconductor \cite{Prozorov2008a,Prozorov2010a}. These experiments indicate a very robust and uniform bulk superconductivity with critical current densities (estimated from $B_z (\vec{r})$ profiles) exceeding 10$^5$ A/cm$^2$. 

\begin{figure}
	\includegraphics[scale=1]{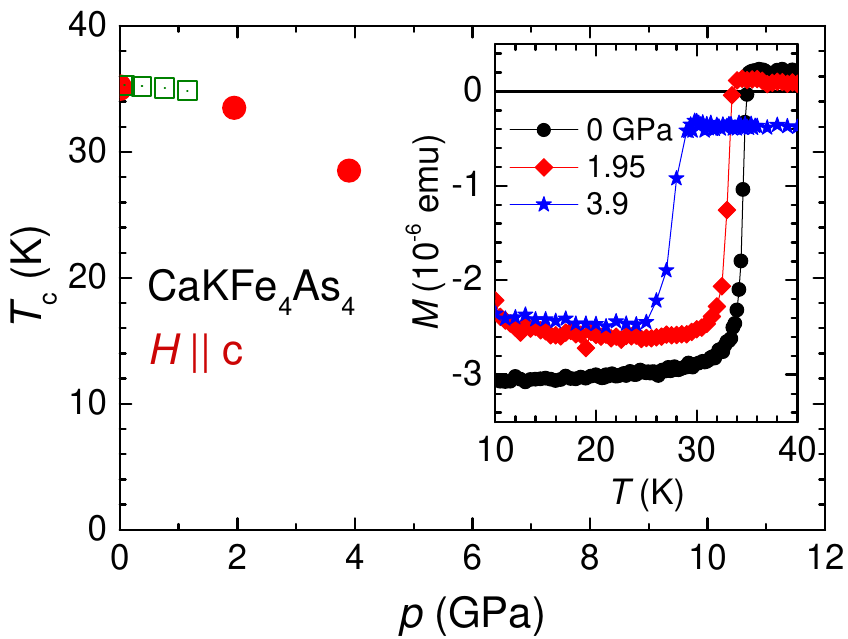}%
	\caption{(Color online) The superconducting critical temperature, $T_c$, of CaKFe$_{4}$As$_{4}$ as a function of applied pressure. Open square symbols from piston cylinder cell and filled symbols from moissanite anvil cell. Inset: $M(T)$ measured in a moissanite anvil cell for $p$~=~0, 1.95, and 3.9\,GPa.
		\label{pressure}}
\end{figure} 

\begin{figure}
	\includegraphics[scale=1]{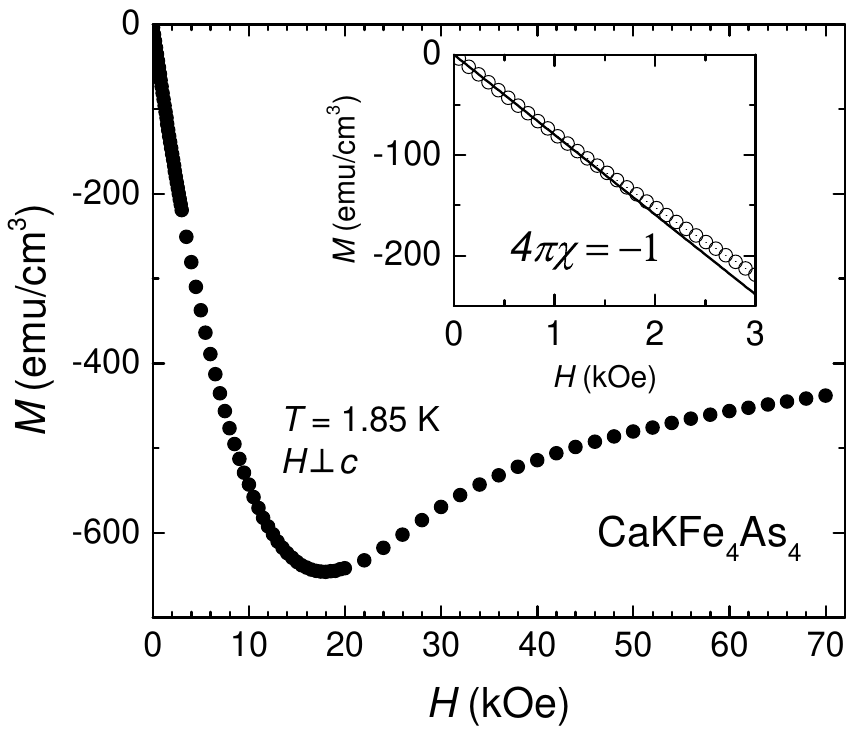}%
	\caption{Magnetization as a function of magnetic field applied perpendicular to the crystallographic \textit{c}-axis of CaKFe$_{4}$As$_{4}$ for $T$~=~1.85\,K. Inset: low field extended view and solid line showing ideal $\chi=-1/4\,\pi$.
		\label{MH}}
\end{figure}

The pressure dependence of $T_c$ was inferred from pressure dependent magnetization measurements. Figure~\ref{pressure} shows that, although there is an initially weak suppression of $T_c$ for $p~\textless$~1.3\,GPa, as pressure is increased, a non-linear, much stronger suppression takes place for 1\,GPa~$\textless~p~\textless$~4\,GPa. By $p$~=~3.9\,GPa, $T_c$ has been suppressed to 28.5\,K. As shown in the inset of Figure \ref{pressure}, the superconducting transition remains sharp up to the $p$~=~3.9\,GPa data point. Higher pressure measurements will be needed to determine the ultimate, critical pressure for superconductivity in this system.

\begin{figure}
	\includegraphics[scale=1]{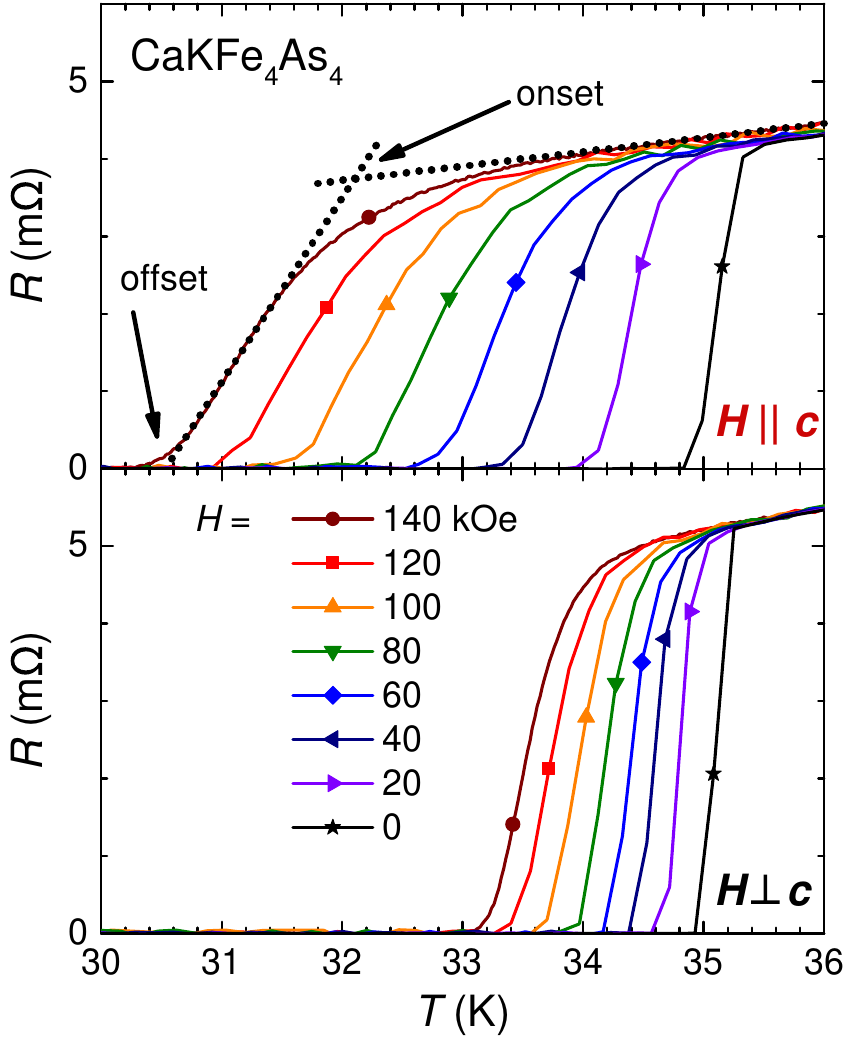}%
	\caption{(Color online) Temperature dependent electrical resistance of CaKFe$_{4}$As$_{4}$ for $H$ applied parallel and perpendicular to the crystallographic \textit{c}-axis for representative fields $ H \leqslant$ 140 kOe. Onset and offset criteria for $T_c$ are shown by dashed lines in the panel.
		\label{RTH}}
\end{figure}

\begin{figure}
	\includegraphics[scale=1]{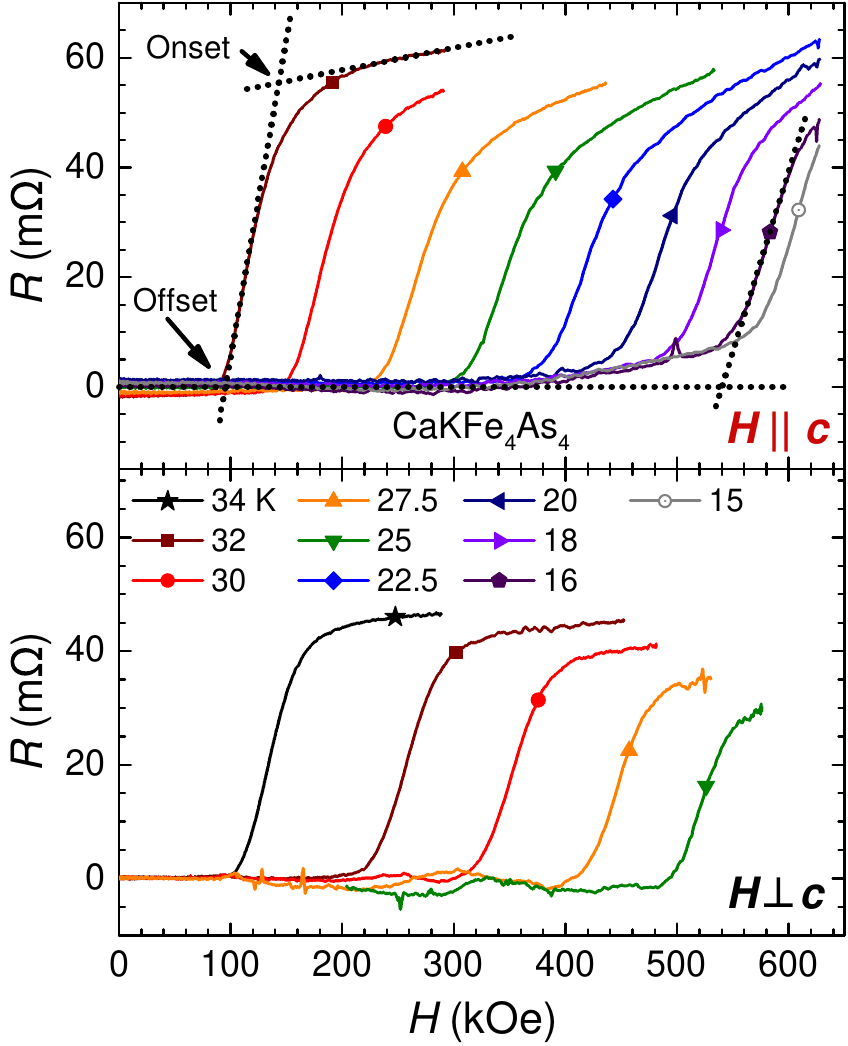}%
	\caption{(Color online) Field-dependent resistance measured in a 630 kOe pulsed magnet at different temperatures with (a) $H\parallel c$ and (b) $H\perp c$. A temperature-independent background signal was subtracted for clarity (see text). Dotted line and arrows indicate different criteria for determining $H_{c\text{2}}$ (see text).
		\label{HighFieldRTH}}
\end{figure}

The superconducting state can also be studied as a function of applied field. Figure~\ref{MH} presents the $M(H)$ isotherm for $T$~=~1.85\,K with $H$ applied within the plane of the crystalline plate (i.e. $H\perp c$). As is shown in the inset, the initial slope is indeed $-1/4\,\pi$ and the measurements start to deviate from this value for $H\lesssim$ 1\,kOe. This sets an upper limit on the low temperature $H_{c1}^{\perp}$ value consistent with the magneto-optical data in Fig.~\ref{RChiCpT}. Figure~\ref{RTH} presents the in-plane, electrical resistance data measured in a QD PPMS using a static magnetic field for $H \leqslant$~140\,kOe for $H\|c$ and $H\perp c$. An example of the criteria used to determine $H_{c2}$($T$) values is shown in the upper panel of Fig.~\ref{RTH}(a). Fig.~\ref{HighFieldRTH} shows the field-dependent resistance measured at different temperatures in a pulsed magnet. A temperature-independent background was subtracted from the signal for clarity. The background is attributed to the displacement of the sample and its wiring by Lorentz force synchronous with lock-in excitation current. The resulting magnetic inductance voltage is a product of field intensity and Lorentz force, leading to a stray background signal proportional to $H^2$. Similar onset and offset criteria were applied to extract the superconducting field values at a given temperature. For $H\| c$ at 15 K, only an offset value could be resolved as shown in Fig.~\ref{HighFieldRTH}.

\begin{figure}
	\includegraphics[scale=1]{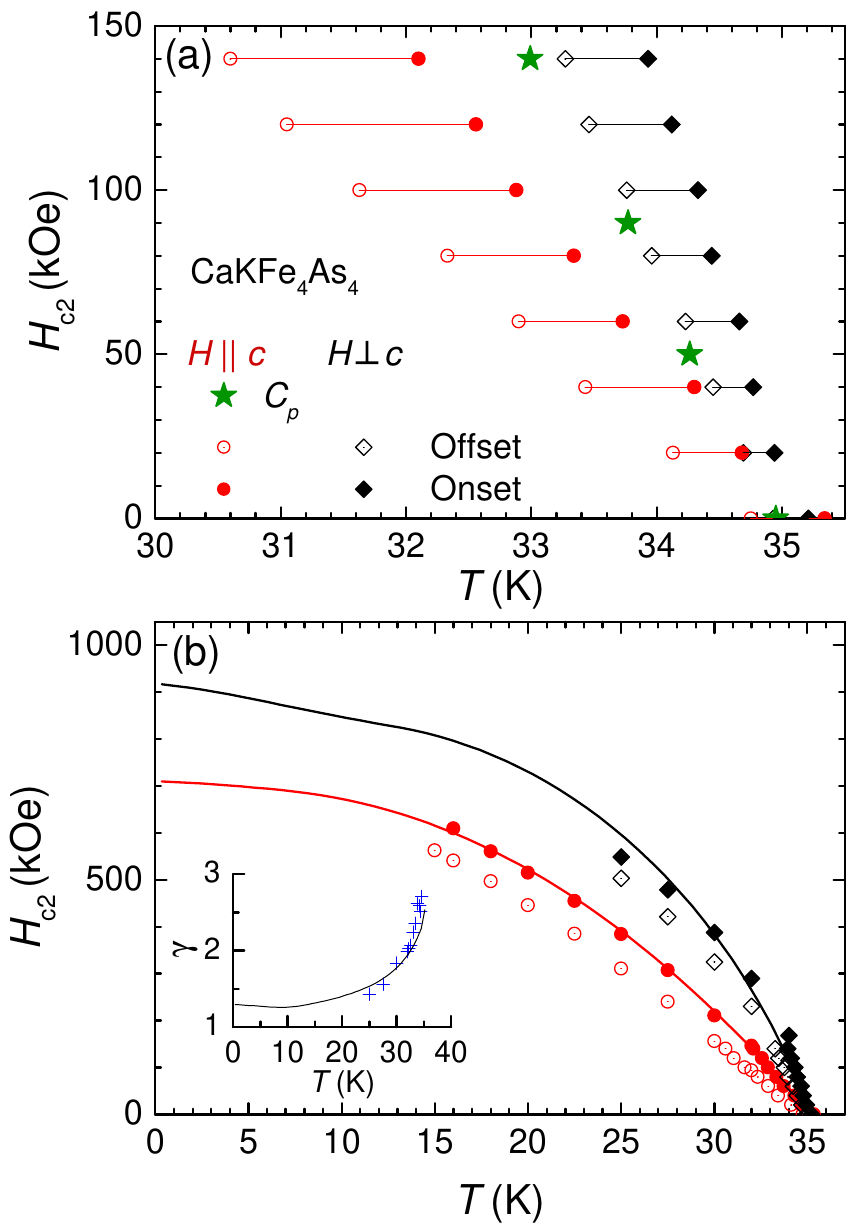}%
	\caption{(Color online)
		(a) Anisotropic $H_{c2}(T)$ data for CaKFe$_{4}$As$_{4}$ inferred from the temperature-dependent electrical resistivity data presented in Fig.~\ref{RTH}. The $H_{c2}(T)$ data for $H\|c$ inferred from temperature and field dependent specific heat measurements (Fig.~\ref{CpTH}), using an equi-entropic mid-point criterion, are also shown. (b) Anisotropic $H_{c\text{2}}$($T$) up to 630 kOe, including the data shown in (a) for field below 140 kOe. Black diamonds represent $H_{c\text{2}}^{\perp}$($T$). Red circles represent $H_{c\text{2}}^{\|}$($T$). Open and filled symbols indicate offset and onset criteria as described in the text. Black and red solid lines in the main figure are theoretically fitted curves to the onset criteria (see text). The inset shows the anisotropic parameter $\gamma(T)=H_{c\text{2}}^{\perp}/H_{c\text{2}}^{\|}$ together with the theoretically fitted curve (black solid line). 
		\label{Hc2T}}
\end{figure}

Fig.~\ref{Hc2T} presents the anisotropic $H_{c2}(T)$ curves for the two directions of applied field. These data make it immediately clear that CaKFe$_{4}$As$_{4}$, like other Fe-based superconductors with comparable $T_c$-values, will have substantial, low temperature $H_{c2}$ values, and will likely have moderate, but not substantial, $H_{c2}(T)$ anisotropy, with the $H\perp c$ manifold being somewhat larger, at least at higher temperatures. Clearly, further measurements for applied fields larger than 630\,kOe will be needed to more fully determine the high field behavior of the superconducting state in CaKFe$_{4}$As$_{4}$.  

\begin{figure}
	\includegraphics[scale=1]{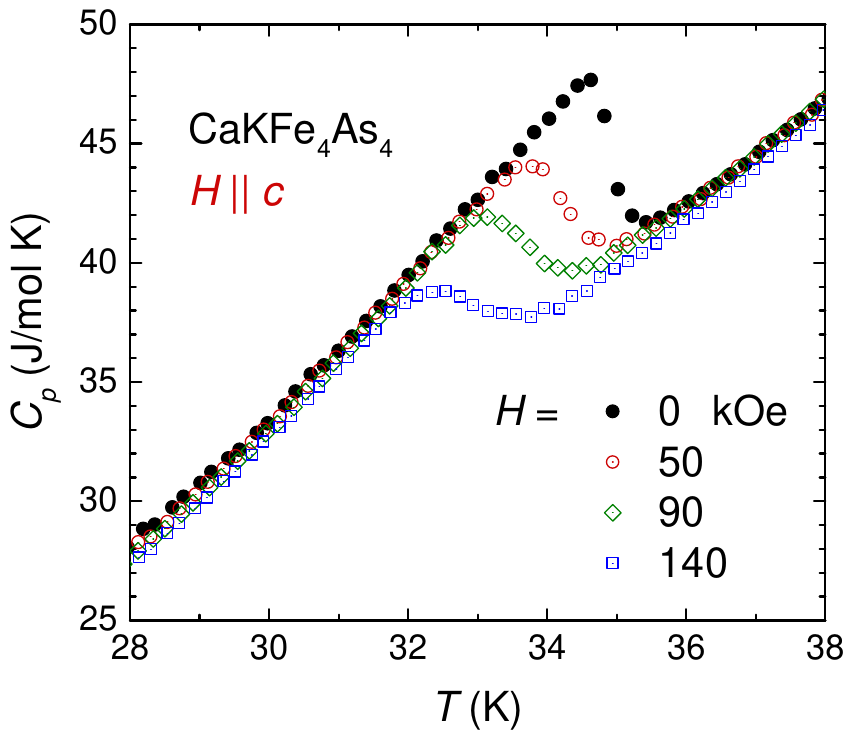}%
	\caption{(Color online) Temperature dependent specific heat data for CaKFe$_{4}$As$_{4}$ taken for $H\|c$~=~0, 50, 90, 140\,kOe. The data for finite $H$ have been normalized to those for $H$~=~0\,kOe in the normal state above $T_c$.
		\label{CpTH}}
\end{figure}

Specific heat data for $H\|c$, $H \leqslant$~140\,kOe were also collected and are shown in Fig.~\ref{CpTH}. $H_{c2}(T)$ data inferred from the specific heat data are also shown in Fig.~\ref{Hc2T}(a). The $H_{c2}(T)$ data inferred from the specific heat data are distinguishably higher than those associated with the electrical resistivity data for the same, $H \|c$, field orientation.  The specific heat inferred $H_{c2}(T)$ manifold is actually closer to that found for $H\perp c$. Given that there was some minor rotation of the specific heat platform (as described in the experimental methods section) it is possible that the difference between the $H_{c2}^{\|}(T)$ manifolds could be associated with a very sharp, or rapid, angular dependence of $H_{c2}(T)$ that has a relative minima for $H\|c$ and even for deviations of 10\textdegree~from $H\|c$ approaches the $H_{c2}(T)$ manifold for $H\perp c$. A second, more likely, explanation for the difference in $H_{c2}(T)$ data for $H \| c$ is that there are significant vortex flow effects that lead to an apparent reduction of the inferred $T_c$ for a given applied field and measurable difference between thermodynamically measured $H_{c2}$ and irreversibility field, $H_{irr}$, inferred from transport measurements. \cite{Welp89}

\section{DISCUSSION}

CaKFe$_{4}$As$_{4}$ is an ordered example of a Fe-based superconductor with a relatively high $T_c$ value and no discernible signature of any other ordering. The data presented in Figs.3-13 are remarkably similar to that measured for optimally- or slightly over-doped (Ba,K)Fe$_2$As$_2$ and Ba(Fe,Co)$_2$As$_2$ compounds. As argued previously\cite{Iyo16}, the unambiguous appearance of $h+k+\ell=$ odd lines, specifically in this case $\ell$ = odd $(00\ell)$ lines, demonstrates a new, ordered structure rather than a (Ca$_{0.5}$K$_{0.5}$)Fe$_{2}$As$_{2}$ solid solution in the body centered $I4/mmm$ structure. The residual resistivity ratio, RRR\,=\,15 is also consistent with an ordered compound, although, by itself, not conclusive. There is no evidence of a structural phase transition down to 6\,K and there is no evidence of a magnetic or electronic phase transition other than superconductivity at $T_c$~=~35\,$\pm$\,0.2\,K. The pressure dependence of $T_c$ is initially very shallow, almost pressure independent up to 1\,GPa, followed by a sharper drop for 1\,GPa~$\textless~p~\textless$~4\,GPa. Based on the response of Ba(Fe$_{1-x}$Co$_{x}$)$_{2}$As$_{2}$ and (Ba$_{1-x}$K$_{x}$)Fe$_{2}$As$_{2}$ across the under-doped, optimally doped, over-doped parts of the phase diagram, \cite{Colombier10,Sergey13} CaKFe$_{4}$As$_{4}$ appears to be near optimal doping.  

The anisotropic $H_{c\text{2}}$($T$) data inferred from the temperature-dependent and field-dependent resistance data, summarized in Fig.~\ref{Hc2T}, reveal multiple features about CaKFe$_4$As$_4$. (1) The values of $H_{c\text{2}}(0)$ both parallel and perpendicular to the $c$-axis extrapolate to the fields well above the single-band BCS paramagnetic limit $H_p [T] = 1.84T_c [K] \simeq 640$ kOe, which is close to the maximum field in our pulse magnet. Thus, Pauli pairbreaking is essential, similar to the majority of other Fe-based superconductors\cite{Gurevich11}. (2) As a result of different temperature dependencies of $H_{c\text{2}}^{\|}(T)$ and $H_{c\text{2}}^{\perp}(T)$ the anisotropy parameter $\gamma(T)=H_{c\text{2}}^{\perp}(T)/H_{c\text{2}}^{\|}(T)$ decreases as $T$ decreases (see lower inset in Fig.~\ref{Hc2T}(b)), consistent with the interplay of orbital and Pauli pairbreaking\cite{Gurevich11}. (3) No crossing of $H_{c\text{2}}^{\|}(T)$ and $H_{c\text{2}}^{\perp}(T)$ was observed for $0<H<630$ kOe, although a possibility that it may happen at higher fields cannot be ruled out.  	
	
The initial ($H$~$\leqslant$~140\,kOe) $H_{c2}(T)$ anisotropy shown in Fig.~\ref{Hc2T} is almost identical to that found for (Ba$_{0.55}$K$_{0.45}$)Fe$_{2}$As$_{2}$\cite{Ni08a,Altarawneh08}. Indeed, based on this and the other similarities to near optimally doped (Ba$_{1-x}$K$_x$)Fe$_2$As$_2$, we can anticipate that the low temperature $H_{c2}$ values will be relatively isotropic and in the 600-800\,kOe range. Taking the onset criteria, a more quantitative analysis of our $H_{c2}(T)$ data shows that the $d H_\text{c2}$/$d T$ values at $T_c$ are -109\,kOe/K and -44\,kOe/K for $H \perp c$ and $H\| c$ respectively. We can use the jump in zero-field specific heat data at $T_c$ (shown in Fig.~\ref{CpTH}) and the Rutgers relation\cite{Rutgers34,Welp89}:
\begin{equation}
	\frac{\Delta C}{T_{c}}=\frac{1}{8\pi\kappa^{2}}\left.\left(\frac{dH_{c2}}{dT}\right)^2\right\vert_{T=T_{c}}
	\label{eq_Rutgers}	
\end{equation} 
\noindent where $\Delta C$~=~8.33$\times10^5$\,erg\,cm$^{-3}$\,K$^{-1}$ and infer values of $\kappa$ to be: 141 and 57 for $H\perp c$ and $H\| c$ respectively.

In the two-band model the slope $dH_{c2}^\|/dT$ at $T_c$  can be expressed in terms of the band parameters as follows \cite{Gurevich11}
 \begin{gather}
\frac{dH_{c2}^{\|}}{dT}=\frac{\phi_0}{2\pi T_c\xi_\perp^2}
\label{sl} \\
\xi_\perp^2=\frac{1}{2}\left[\left(1+\frac{\lambda_-}{\lambda_0}\right)\xi_1^2+\left(1-\frac{\lambda_-}{\lambda_0}\right)\xi_2^2\right],
\label{tb}
\end{gather}   
where the effective Ginzburg-Landau coherence length $\xi_\perp$ determines the magnitude of the temperature-dependent $\xi_\perp(T)=\xi_\perp \tau^{-1/2}$ near $T_c$, $\tau = 1-T/T_c$, $\phi_0$ is the magnetic flux quantum,
$\lambda_0=(\lambda_-^2+4\lambda_{12}\lambda_{21})^{1/2}$, $\lambda_-=\lambda_{11}-\lambda_{22}$, $\lambda_{11}$ and $\lambda_{22}$ are dimensionless pairing constants in bands 1 and 2, and $\lambda_{12}$ and $\lambda_{21}$ are interband pairing constants. Eqs. (\ref{sl}) and (\ref{tb}) are applicable for both clean and dirty limits. In the clean limit, the partial coherence lengths $\xi_1=(7\zeta(3)/3)^{1/2}\hbar v_1/4\pi k_BT_c$ and $\xi_2=(7\zeta(3)/3)^{1/2}\hbar v_2/4\pi k_BT_c$ are proportional to the in-plane Fermi velocities $v_1$ and $v_2$ in bands 1 and 2. If the $s^\pm$ pairing in Fe-based superconductors  is dominated by interband coupling \cite{Hirschfeld11}, Eq. (\ref{tb}) yields $\xi_\perp^2\to (\xi_1^2+\xi_2^2)/2$ at $\lambda_-\ll \lambda_0$.     

If both bands have the same mass anisotropy parameter $\epsilon = m_{\perp}/m_{\|}<1$, the values of $\xi_{\perp}$ in the $ab$ plane and $\xi_{\|}$ along the $c$-axis, can be estimated using the anisotropic scaling relations $|dH_{c\text{2}}^{\|}/dT|=\phi_0/2\pi\xi_{\perp}^2T_c$ and $|dH_{c\text{2}}^{\perp}/dT|=\phi_0/2\pi\xi_{\perp}\xi_{\|}T_c$. Hence, we obtain $\xi_{\perp}\simeq 14.3\, \text{\AA}$ and $\xi_{\|}\simeq 5.8\, \text{\AA}$. These coherence lengths are of the order of the lattice parameters shown in Fig.\ref{latticeT} with the value $\xi_{\|}$ being about half the unit cell height along the c-axis. To see the ratio of $\xi_\perp$ to the mean free path $l$, we estimate $l$ using a single-band anisotropic Drude formula $l=\hbar(3\pi^2n\sqrt{\epsilon})^{1/3}/ne^2\rho_n$. Taking $\rho_n= 20\, \mu\Omega$ cm at $T=T_c$ which is 15 times smaller that $\rho_n$ at 300 K, $n=7.4\times 10^{21}$ cm$^{-3}$ and $\epsilon = m_{\perp}/m_{\|} = (\xi_{\|}/\xi_{\perp})^2 = 1/6$, we obtain $l\approx 125\,\text{\AA} $. This rough estimate suggests that our sample is in a  clean limit with $\xi_\perp\ll l$.

To gain further insight into the behavior of $H_{c\text{2}}(T)$, we fitted the experimental data using a two-band theory which takes into account both orbital and Pauli pairbreaking in the clean limit for two ellipsoidal Fermi surfaces. In this case the equation for $H_{c\text{2}}^{\|}$ is given by \cite{Gurevich11},
\begin{gather}
    a_1G_1+a_2G_2+G_1G_2=0,
\label{mgh} \\
    G_1=\ln t+2e^{q^{2}}\operatorname{Re}\sum_{n=0}^{\infty}\int_{q}^{\infty}due^{-u^{2}}\times
\nonumber \\
    \left[\frac{u}{n+1/2}-\frac{t}{\sqrt{b}}\tan^{-1}\left(
    \frac{u\sqrt{b}}{t(n+1/2)+i\alpha b}\right)\right].
\label{U1}
\end{gather}
Here $a_1=(\lambda_0 +\lambda_{-})/2w$, $a_2=(\lambda_0-\lambda_{-})/2w$, and $w=\lambda_{11}\lambda_{22}-\lambda_{12}\lambda_{21}$, $t=T/T_c$. The function $G_2$ 
is obtained by replacing $\sqrt{b}\to\sqrt{\eta b}$, $q\to q\sqrt{s}$, $g_1\to g_2$ in $G_1$, where
\begin{gather}
    b=\frac{\hbar^{2}v^{2}_1 H_{c2} }{8\pi\phi_{0}k_B^2T_{c}^{2}},\qquad\alpha=\frac{4\mu \phi_{0}k_BT_{c}}{\hbar^{2}v^2_1},
    \label{parm1} \\
    q^{2}=Q^{2}\phi_{0}\epsilon_1/2\pi H_{c2}, \qquad \eta = v_2^2/v_1^2, \qquad
    s=\epsilon_2/\epsilon_1.
    \label{parm2}
\end{gather}
Here $Q$ is the wave vector of the Fulde-Ferrell-Larkin-Ovchinnikov (FFLO) modulations of the order parameter, $v_j$ is the in-plane Fermi velocity in band $j={1,2}$, $\epsilon_{j}=m_{j}^{\perp}/m_{j}^{\|}$ is the mass anisotropy ratio, $\mu$ is the magnetic moment of a quasiparticle, $\alpha\approx 1.8\alpha_M$, $\alpha_M=H_{c\text{2}}^{orb}/\sqrt{2}H_p$ is the Maki paramagnetic parameter. Eqs. (\ref{mgh})-(\ref{U1}) do not not take into account spin-orbit effects, and the renormalized values of $T_c$, $v_1$, $v_2$ and $\mu$ include corrections coming from the Fermi liquid and strong coupling effects. If $\textbf{H}$ is applied along the symmetry axis, $\textbf{Q}$ is parallel to $\textbf{H}$ and the magnitude of $Q$ is determined by the condition $\partial H_{c\text{2}}/\partial Q=0$ of maximum $H_{c\text{2}}$. If $\epsilon_1=\epsilon_2=\epsilon$, the anisotropic $H_{c\text{2}}$ can be written in the scaling form 
 $$
 H_{c\text{2}}^{\|}(T) = H_0b(t,\eta,\alpha),\quad H_{c\text{2}}^{\perp}(T) = \frac{H_0}{\sqrt{\epsilon}}b\left(t,\eta,\frac{\alpha}{\sqrt{\epsilon}}\right), 
 $$
 where $H_0=8\pi\phi_0k_B^2T_c^2/\hbar^2v_1^2$ and $b$ is a solution of Eq. (\ref{mgh}). The fit of the measured $H_{c\text{2}}(T)$ to Eq. (\ref{mgh}) for $s_{\pm}$ pairing with $\lambda_{11}=\lambda_{22}=0$, $\lambda_{12}\lambda_{21}=0.25$, $\eta = 0.2$, $\alpha=0.5$, and $\epsilon = 1/6$ is shown in Fig.~\ref{Hc2T} where $H_0$ was adjusted to fit the magnitude of $H_{c\text{2}}^{\|}(T)$. The value of $\alpha$ is consistent with those which have been used previously to describe $H_{c\text{2}}(T)$ of Ba$_{1-x}$K$_x$As$_2$Fe$_2$ \cite{Tarantini11}.  
 
The fit shows that the upper critical fields at $T=0$ extrapolate to $H_{c\text{2}}^{\|}(0) \approx 710$ kOe and $H_{c\text{2}}^{\perp}(0) \approx 920$ kOe, the shape of $H_{c\text{2}}^{\|}(T)$ being mostly determined by orbital effects moderately affected by the Pauli pairbreaking. By contrast, the shape of $H_{c\text{2}}^{\perp}(T)$ is consistent with the essential Pauli pairbreaking in both bands, because of large respective Maki parameters $\alpha_1^{\perp} = \alpha/\sqrt{\epsilon}$ and $\alpha_2^{\perp} = \alpha/\eta\sqrt{\epsilon}$. As a result, the anisotropy parameter $\gamma(T)=H_{c2}^\perp(T)/H_{c2}^\|(T)$  decreases with $T$, which reflects different temperature dependencies of the orbitally-limited and Pauli-limited upper critical fields. 

It should be mentioned that in the available field range $0<H<630$~kOe where the $H_{c2}$ data were obtained, the fit is not very sensitive to the particular values of the pairing constants and the band asymmetry parameter $\eta$, yet it suggests the possibility of a FFLO state for $T < 13$ K and for higher fields $H$ parallel to the $ab$ planes. In fact, the data shown in Fig. \ref{Hc2T} could be fitted equally well with a single-band model in which $H_{c2}(T)$ is defined by the equation, $G_1(b)=0$. More definite conclusions about multiband orbital effects and FFLO states could be made by analyzing low-temperature parts of the $H_{c\text{2}}^{\|}(T)$ and $H_{c\text{2}}^{\perp}(T)$, which would require even higher fields $H>630$ kOe. This distinguishes CaKFe$_4$As$_4$ from other ordered stoichiometric Fe-based superconducting compounds like LiFeAs for which the entire anisotropic $H_{c\text{2}}(T)$ has been measured \cite{Cho11}.   

Further insights into the magneto-transport behavior of CaKFe$_4$As$_4$ can be inferred from the fact that the resistance transition curves $R(T)$ shown in Figs. \ref{RTH} broaden as $H$ increases. This indicates a possible effect of thermal fluctuations of vortices similar to that has been extensively studied in high-$T_c$ \cite{Blatter94}. Broadening of the superconducting transition in CaKFe$_4$As$_4$ under magnetic field is also clearly seen in the behavior of the specific heat shown in Fig. \ref{CpTH}. 

At $H=0$ thermal fluctuations can be quantified by the Ginzburg number $Gi=0.5(2\pi \mu_0 k_BT_c\lambda_0^2/\phi_0\xi_{\|})^2$ expressed in terms of $\xi_\|$ and the London penetration depth $\lambda_0$ at $H\|c$ and $T=0$. Using the values of $\lambda_0=133$ nm\cite{Cho16}, $\xi_{\|}=0.6$ nm, and $T_c=35$ K, we obtain that CaKFe$_4$As$_4$ would have $Gi\simeq 4\cdot 10^{-4}$ of the same order of magnitude as $Gi$ for BaFe$_2$As$_2$-based compounds, but smaller than $Gi\sim 10^{-2}$ for YBa$_2$Cu$_3$O$_{7-x}$ \cite{Putti10,Gurevich14}. The irreversibility field $H_p(T)$ associated with the offset point of $R(T,H)=0$ in Fig.~\ref{RTH} can be qualitatively evaluated in terms of melting and thermal depinning of vortex structure. For instance, the melting field $H_m$ of the ideal vortex lattice in a uniaxial superconductor at $H\|c$ is defined by the equation $h_m/(1-h_m)^3=(1-t)t_0^2/t^2$, where $h_m=H_m/H_{c\text{2}}$, $t_0=\pi c_L^2/Gi^{1/2}$ and $c_L=0.15-0.17$ is the Lindemann number \cite{Blatter94}. For weak thermal fluctuations, $H_{c\text{2}}-H_m \ll H_{c\text{2}}$, the above equation for $h_m$ yields

\begin{equation}
H_{c\text{2}}(T)-H_m(T)\simeq H_{c\text{2}}(0)\left(\frac{Gi}{\pi^2 c_L^4}\right)^{1/3}\!\!\left(1-\frac{T}{T_c}\right)^{2/3}
\label{hm}
\end{equation}   
Taking $c_L=0.15$ and $Gi=4\cdot10^{-4}$ in Eq. (\ref{hm}) gives $(Gi/\pi^2c_L^4)^{1/3}\approx 0.43$, which shows that thermal fluctuations in CaKFe$_4$As$_4$ are not weak, as also characteristic of the majority of Fe-based superconductors which are intermediate between the conventional low-$T_c$ superconductors in which vortex fluctuations are negligible and high-$T_c$ cuprates in which the behavior of vortex matter at 77K is controlled by thermal fluctuations \cite{Putti10,Gurevich14}. Yet the width of the critical fluctuation region $T_c-T \lesssim Gi T_c\sim 0.014$ K is much smaller that the observed width of the sharp resistive transition $\Delta T \simeq 0.4$ K at $H=0$ shown in Fig.~\ref{RChiCpT}, as well as the width of the step in specific heat in zero field. This suggests that, in addition to thermal fluctuations of the order parameter, the resistive transition at zero field can be broadened by extrinsic factors such as weak materials' inhomogeneities in $T_c$. As $H$ increases, the field-induced broadening of the resistive transition becomes more pronounced, structural defects and inhomogeneities in $T_c$ affecting both the thermally-activated flux flow resistance \cite{Blatter94} and the vortex melting field \cite{Mikitik03}.   

\begin{figure}
	\includegraphics[scale=1]{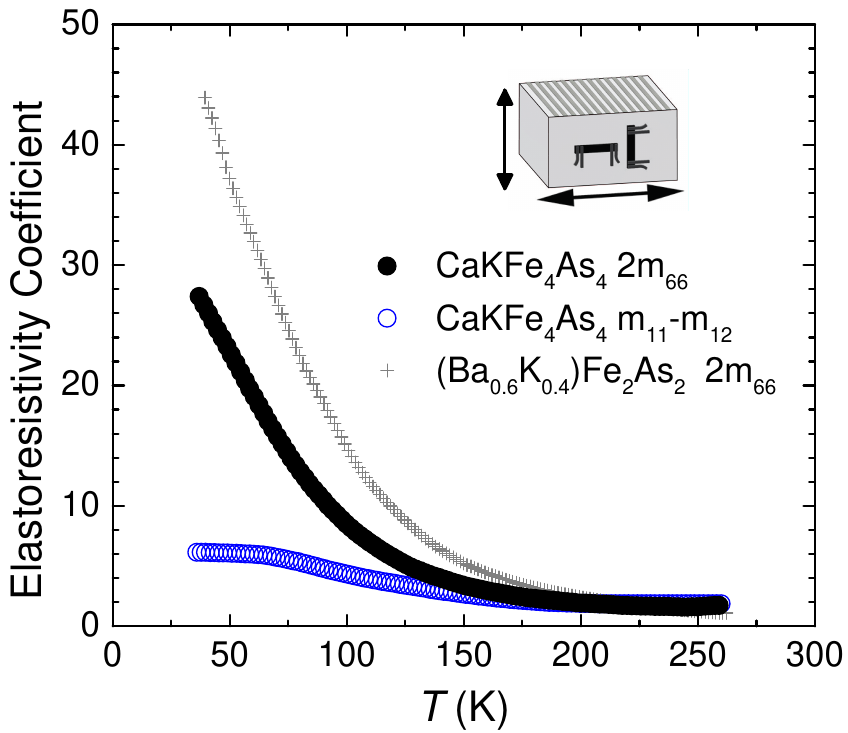}%
	\caption{(Color online) Elastoresistivity coefficients of $2m_{66}$ and $m_{11}-m_{12}$ of CaKFe$_{4}$As$_{4}$ (open and filled circles) measured using crossed samples glued to a piezostack, shown schematically in the right inset. The $2m_{66}$ coefficient data of optimally doped K-doped BaFe$_{2}$As$_{2}$ (grey +'s) from Ref.~\onlinecite{Kuo16} are plotted for comparison.
		\label{elastoresistivity}}
\end{figure}

To further explore the similarity between CaKFe$_4$As$_4$ and near-optimally doped (Ba$_{1-x}$K$_x$)Fe$_2$As$_2$ we determined the elastoresistivity coefficients 2$m_{66}$ and $m_{11}-m_{12}$ of CaKFe$_4$As$_4$ using a piezo-stack-based setup; these data are presented in Fig.~\ref{elastoresistivity}. For comparison the 2$m_{66}$ coefficient data of near-optimally doped (Ba$_{1-x}$K$_x$)Fe$_2$As$_2$ from Ref.~\onlinecite{Kuo16} are also shown. The elastoresistivity coefficients are defined in the tetragonal unit cell. 2$m_{66}$ measures the size of the resistivity anisotropy along the Fe-Fe bonds (the diagonals of the tetragonal unit cell) $\rho_{[110]}-\rho_{[1\bar{1}0]}$ induced by the corresponding shear strain $\varepsilon_{[110]}- \varepsilon_{[1\bar{1}0]}$,
\begin{equation}
	2m_{66} = \frac{1}{\rho}~\frac{d\left(\rho_{[110]}-\rho_{[1\bar{1}0]}\right)}{d\left(\varepsilon_{[110]}- \varepsilon_{[1\bar{1}0]}\right)}.
\end{equation}
In typical Fe-based superconductors, $m_{66}$ is closely related to the nematic susceptibility $\chi_{\mathrm{nem}}$. It is expected to diverge on approaching the nematic (tetragonal-to-orthorhombic) transition in under-doped samples \cite{Chu12}, in which the Fe-Fe bonds become inequivalent. Similarly to the optimally K-doped BaFe$_2$As$_2$, the 2$m_{66}$ coefficient of CaKFe$_4$As$_4$ indeed rises strongly with decreasing temperature, indicating proximity to a nematic transition. Note that, despite its strong increase at low temperatures, 2$m_{66}$ does not show Curie-Weiss type divergence \cite{Chu12, Kuo16} for either compound. In contrast, the elastoresistivity mode, $m_{11}-m_{12}$, shows only a weak temperature dependence in CaKFe$_4$As$_4$.  It is related to the sensitivity of the resistivity anisotropy between the two tetragonal in-plane axes to the corresponding shear strain
\begin{equation}
	m_{11}-m_{12}= \frac{1}{\rho}~\frac{d\left(\rho_{[100]}-\rho_{[010]}\right)}{d\left(\varepsilon_{[100]}-\varepsilon_{[010]}\right)}.
\end{equation}
This mode does not directly couple to the nematic order parameter of typical under-doped Fe-based systems. All in all, the elastoresistivity data of CaKFe$_4$As$_4$ indicates that it is close to a nematic structural instability, similarly to other optimally-doped Fe-based superconductors. \cite{Kuo16}

\begin{figure}
	\includegraphics[scale=1]{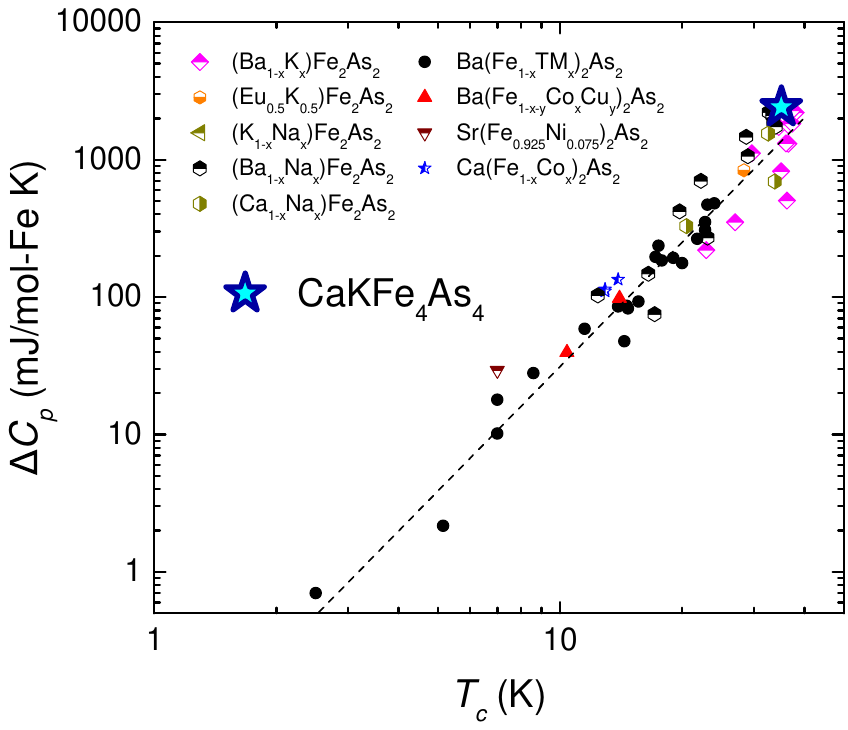}%
	\caption{(Color online) Log-log plot of $\Delta C_p$ jump at $T_c$ versus $T_c$ (BNC plot [Refs.~\onlinecite{Stewart11,Ran12,Sergey09}]). Note that data for (Ba$_{1-x}$K$_{x}$)Fe$_{2}$As$_{2}$ are plotted for $x~\textless~0.8$. For Ba(Fe$_{1-x}$TM$_x$)$_2$As$_2$, TM = Ni, Co, Rh, Pd, Pt [Ref. \onlinecite{Ran12}]. Dashed line has a slope corresponding to $\Delta$C$_p \sim T_c^3$ and is a guide for the eyes.
		\label{BNC}}
\end{figure}

CaKFe$_{4}$As$_{4}$ can also be put in context of other AeFe$_2$As$_2$-based (Ae = Ba, Sr, Ca) superconductors by placing it on a $\Delta$C$_p$ versus T$_c$, or BNC, scaling \cite{Stewart11,Sergey09} plot (Fig.~\ref{BNC}). The jump in specific heat of CaKFe$_{4}$As$_{4}$ is sharp and well defined (perhaps due, in part, to its fully ordered nature) and, combined with its $T_c$ value places CaKFe$_{4}$As$_{4}$ at the extreme, near optimally doped end of the BNC data set for AeFe$_{2}$As$_{2}$ systems.

\section{CONCLUSIONS}
We have synthesized single phase, single crystalline samples of CaKFe$_{4}$As$_{4}$ and measured temperature dependent unit cell  dimensions, temperature and field dependent specific heat as well as thermoelectric power, Hall effect, elastoresistivity, and anisotropic temperature and field dependent magnetization and electrical resistivity. There is no indication of any phase transition, other than superconductivity with $T_c$~=~35.0\,$\pm$\,0.2\,K, taking place in this compound for 1.8\,K~$\leq$~$T$~$\leq $~300\,K. The temperature dependence of our thermodynamic and transport measurements, the resistive anisotropy, the pressure dependence of $T_c$, and the anisotropy and size of $H_{c2}$(T) are consistent with near optimally doped members of the (Ba$_{1-x}$K$_{x}$)Fe$_{2}$As$_{2}$ series. In addition CaKFe$_{4}$As$_{4}$ falls directly onto the BNC scaling plot at the near optimal end of the AeFe$_2$As$_2$ structure manifold. All of these data indicate that stoichiometric CaKFe$_{4}$As$_{4}$ is intrinsically close to what is referred to as "optimal-doped" on a generalized, Fe-based superconductor, phase diagram.

\begin{acknowledgements}

We would like to thank J. Betts, M. Jaime, R. McDonald, B. Ramshaw and M. Chan for useful discussions and experimental assistance. This work was supported by the U.S. Department of Energy, Office of Basic Energy Science, Division of Materials Sciences and Engineering. The research was performed at the Ames Laboratory. Ames Laboratory is operated for the U.S. Department of Energy by Iowa State University under Contract No. DE-AC02-07CH11358. In addition, G. D., N. H. J., and W. M. were supported by the Gordon and Betty Moore Foundation’s EPiQS Initiative through Grant GBMF4411. V. T. is supported by Ames Laboratory's laboratory-directed research and development (LDRD) funding for magnetization measurements under pressure. We are grateful to D. S. Robinson for support during the high energy x-ray experiments. This research used resources of the Advanced Photon Source, a US Department of Energy (DOE) Office of Science User Facility operated for the DOE Office of Science by Argonne National Laboratory under Contract No. DE-AC02- 06CH11357. The NHMFL Pulsed Field Facility is supported by the National Science Foundation, the Department of Energy, and the State of Florida through NSF Cooperative Grant No. DMR-1157490 and by U.S. DOE BES “Science at 100T” project.

\end{acknowledgements}

\bibliographystyle{apsrev4-1}

\begin{thebibliography}{56}%
\makeatletter
\providecommand \@ifxundefined [1]{%
 \@ifx{#1\undefined}
}%
\providecommand \@ifnum [1]{%
 \ifnum #1\expandafter \@firstoftwo
 \else \expandafter \@secondoftwo
 \fi
}%
\providecommand \@ifx [1]{%
 \ifx #1\expandafter \@firstoftwo
 \else \expandafter \@secondoftwo
 \fi
}%
\providecommand \natexlab [1]{#1}%
\providecommand \enquote  [1]{``#1''}%
\providecommand \bibnamefont  [1]{#1}%
\providecommand \bibfnamefont [1]{#1}%
\providecommand \citenamefont [1]{#1}%
\providecommand \href@noop [0]{\@secondoftwo}%
\providecommand \href [0]{\begingroup \@sanitize@url \@href}%
\providecommand \@href[1]{\@@startlink{#1}\@@href}%
\providecommand \@@href[1]{\endgroup#1\@@endlink}%
\providecommand \@sanitize@url [0]{\catcode `\\12\catcode `\$12\catcode
  `\&12\catcode `\#12\catcode `\^12\catcode `\_12\catcode `\%12\relax}%
\providecommand \@@startlink[1]{}%
\providecommand \@@endlink[0]{}%
\providecommand \url  [0]{\begingroup\@sanitize@url \@url }%
\providecommand \@url [1]{\endgroup\@href {#1}{\urlprefix }}%
\providecommand \urlprefix  [0]{URL }%
\providecommand \Eprint [0]{\href }%
\providecommand \doibase [0]{http://dx.doi.org/}%
\providecommand \selectlanguage [0]{\@gobble}%
\providecommand \bibinfo  [0]{\@secondoftwo}%
\providecommand \bibfield  [0]{\@secondoftwo}%
\providecommand \translation [1]{[#1]}%
\providecommand \BibitemOpen [0]{}%
\providecommand \bibitemStop [0]{}%
\providecommand \bibitemNoStop [0]{.\EOS\space}%
\providecommand \EOS [0]{\spacefactor3000\relax}%
\providecommand \BibitemShut  [1]{\csname bibitem#1\endcsname}%
\let\auto@bib@innerbib\@empty
\bibitem [{\citenamefont {Canfield}\ and\ \citenamefont
  {Bud'ko}(2010)}]{Canfield10}%
  \BibitemOpen
  \bibfield  {author} {\bibinfo {author} {\bibfnamefont {P.~C.}\ \bibnamefont
  {Canfield}}\ and\ \bibinfo {author} {\bibfnamefont {S.~L.}\ \bibnamefont
  {Bud'ko}},\ }\href {\doibase 10.1146/annurev-conmatphys-070909-104041}
  {\bibfield  {journal} {\bibinfo  {journal} {Annu. Rev. Cond. Matt. Phys.}\
  }\textbf {\bibinfo {volume} {1}},\ \bibinfo {pages} {27} (\bibinfo {year}
  {2010})}\BibitemShut {NoStop}%
\bibitem [{\citenamefont {Rotter}\ \emph
  {et~al.}(2008{\natexlab{a}})\citenamefont {Rotter}, \citenamefont {Tegel},\
  and\ \citenamefont {Johrendt}}]{Rotter08b}%
  \BibitemOpen
  \bibfield  {author} {\bibinfo {author} {\bibfnamefont {M.}~\bibnamefont
  {Rotter}}, \bibinfo {author} {\bibfnamefont {M.}~\bibnamefont {Tegel}}, \
  and\ \bibinfo {author} {\bibfnamefont {D.}~\bibnamefont {Johrendt}},\ }\href
  {\doibase 10.1103/PhysRevLett.101.107006} {\bibfield  {journal} {\bibinfo
  {journal} {Phys. Rev. Lett.}\ }\textbf {\bibinfo {volume} {101}},\ \bibinfo
  {pages} {107006} (\bibinfo {year} {2008}{\natexlab{a}})}\BibitemShut
  {NoStop}%
\bibitem [{\citenamefont {Ni}\ \emph {et~al.}(2008{\natexlab{a}})\citenamefont
  {Ni}, \citenamefont {Bud'ko}, \citenamefont {Kreyssig}, \citenamefont
  {Nandi}, \citenamefont {Rustan}, \citenamefont {Goldman}, \citenamefont
  {Gupta}, \citenamefont {Corbett}, \citenamefont {Kracher},\ and\
  \citenamefont {Canfield}}]{Ni08a}%
  \BibitemOpen
  \bibfield  {author} {\bibinfo {author} {\bibfnamefont {N.}~\bibnamefont
  {Ni}}, \bibinfo {author} {\bibfnamefont {S.~L.}\ \bibnamefont {Bud'ko}},
  \bibinfo {author} {\bibfnamefont {A.}~\bibnamefont {Kreyssig}}, \bibinfo
  {author} {\bibfnamefont {S.}~\bibnamefont {Nandi}}, \bibinfo {author}
  {\bibfnamefont {G.~E.}\ \bibnamefont {Rustan}}, \bibinfo {author}
  {\bibfnamefont {A.~I.}\ \bibnamefont {Goldman}}, \bibinfo {author}
  {\bibfnamefont {S.}~\bibnamefont {Gupta}}, \bibinfo {author} {\bibfnamefont
  {J.~D.}\ \bibnamefont {Corbett}}, \bibinfo {author} {\bibfnamefont
  {A.}~\bibnamefont {Kracher}}, \ and\ \bibinfo {author} {\bibfnamefont
  {P.~C.}\ \bibnamefont {Canfield}},\ }\href {\doibase
  10.1103/PhysRevB.78.014507} {\bibfield  {journal} {\bibinfo  {journal} {Phys.
  Rev. B}\ }\textbf {\bibinfo {volume} {78}},\ \bibinfo {pages} {014507}
  (\bibinfo {year} {2008}{\natexlab{a}})}\BibitemShut {NoStop}%
\bibitem [{\citenamefont {Sefat}\ \emph {et~al.}(2008)\citenamefont {Sefat},
  \citenamefont {Jin}, \citenamefont {McGuire}, \citenamefont {Sales},
  \citenamefont {Singh},\ and\ \citenamefont {Mandrus}}]{Sefat08}%
  \BibitemOpen
  \bibfield  {author} {\bibinfo {author} {\bibfnamefont {A.~S.}\ \bibnamefont
  {Sefat}}, \bibinfo {author} {\bibfnamefont {R.}~\bibnamefont {Jin}}, \bibinfo
  {author} {\bibfnamefont {M.~A.}\ \bibnamefont {McGuire}}, \bibinfo {author}
  {\bibfnamefont {B.~C.}\ \bibnamefont {Sales}}, \bibinfo {author}
  {\bibfnamefont {D.~J.}\ \bibnamefont {Singh}}, \ and\ \bibinfo {author}
  {\bibfnamefont {D.}~\bibnamefont {Mandrus}},\ }\href {\doibase
  10.1103/PhysRevLett.101.117004} {\bibfield  {journal} {\bibinfo  {journal}
  {Phys. Rev. Lett.}\ }\textbf {\bibinfo {volume} {101}},\ \bibinfo {pages}
  {117004} (\bibinfo {year} {2008})}\BibitemShut {NoStop}%
\bibitem [{\citenamefont {Johnston}(2010)}]{Johnston10}%
  \BibitemOpen
  \bibfield  {author} {\bibinfo {author} {\bibfnamefont {D.~C.}\ \bibnamefont
  {Johnston}},\ }\href {\doibase 10.1080/00018732.2010.513480} {\bibfield
  {journal} {\bibinfo  {journal} {Adv. Phys.}\ }\textbf {\bibinfo {volume}
  {59}},\ \bibinfo {pages} {803} (\bibinfo {year} {2010})}\BibitemShut
  {NoStop}%
\bibitem [{\citenamefont {Stewart}(2011)}]{Stewart11}%
  \BibitemOpen
  \bibfield  {author} {\bibinfo {author} {\bibfnamefont {G.~R.}\ \bibnamefont
  {Stewart}},\ }\href {\doibase 10.1103/RevModPhys.83.1589} {\bibfield
  {journal} {\bibinfo  {journal} {Rev. Mod. Phys.}\ }\textbf {\bibinfo {volume}
  {83}},\ \bibinfo {pages} {1589} (\bibinfo {year} {2011})}\BibitemShut
  {NoStop}%
\bibitem [{\citenamefont {Ni}\ \emph {et~al.}(2008{\natexlab{b}})\citenamefont
  {Ni}, \citenamefont {Nandi}, \citenamefont {Kreyssig}, \citenamefont
  {Goldman}, \citenamefont {Mun}, \citenamefont {Bud'ko},\ and\ \citenamefont
  {Canfield}}]{Ni08b}%
  \BibitemOpen
  \bibfield  {author} {\bibinfo {author} {\bibfnamefont {N.}~\bibnamefont
  {Ni}}, \bibinfo {author} {\bibfnamefont {S.}~\bibnamefont {Nandi}}, \bibinfo
  {author} {\bibfnamefont {A.}~\bibnamefont {Kreyssig}}, \bibinfo {author}
  {\bibfnamefont {A.~I.}\ \bibnamefont {Goldman}}, \bibinfo {author}
  {\bibfnamefont {E.~D.}\ \bibnamefont {Mun}}, \bibinfo {author} {\bibfnamefont
  {S.~L.}\ \bibnamefont {Bud'ko}}, \ and\ \bibinfo {author} {\bibfnamefont
  {P.~C.}\ \bibnamefont {Canfield}},\ }\href {\doibase
  10.1103/PhysRevB.78.014523} {\bibfield  {journal} {\bibinfo  {journal} {Phys.
  Rev. B}\ }\textbf {\bibinfo {volume} {78}},\ \bibinfo {pages} {014523}
  (\bibinfo {year} {2008}{\natexlab{b}})}\BibitemShut {NoStop}%
\bibitem [{\citenamefont {Kamihara}\ \emph {et~al.}(2008)\citenamefont
  {Kamihara}, \citenamefont {Watanabe}, \citenamefont {Hirano},\ and\
  \citenamefont {Hosono}}]{Kamihara08}%
  \BibitemOpen
  \bibfield  {author} {\bibinfo {author} {\bibfnamefont {Y.}~\bibnamefont
  {Kamihara}}, \bibinfo {author} {\bibfnamefont {T.}~\bibnamefont {Watanabe}},
  \bibinfo {author} {\bibfnamefont {M.}~\bibnamefont {Hirano}}, \ and\ \bibinfo
  {author} {\bibfnamefont {H.}~\bibnamefont {Hosono}},\ }\href {\doibase
  10.1021/ja800073m} {\bibfield  {journal} {\bibinfo  {journal} {J. Am. Chem.
  Soc.}\ }\textbf {\bibinfo {volume} {130}},\ \bibinfo {pages} {3296} (\bibinfo
  {year} {2008})}\BibitemShut {NoStop}%
\bibitem [{\citenamefont {Canfield}\ \emph {et~al.}(2009)\citenamefont
  {Canfield}, \citenamefont {Bud'ko}, \citenamefont {Ni}, \citenamefont
  {Kreyssig}, \citenamefont {Goldman}, \citenamefont {McQueeney}, \citenamefont
  {Torikachvili}, \citenamefont {Argyriou}, \citenamefont {Luke},\ and\
  \citenamefont {Yu}}]{Canfield09}%
  \BibitemOpen
  \bibfield  {author} {\bibinfo {author} {\bibfnamefont {P.}~\bibnamefont
  {Canfield}}, \bibinfo {author} {\bibfnamefont {S.}~\bibnamefont {Bud'ko}},
  \bibinfo {author} {\bibfnamefont {N.}~\bibnamefont {Ni}}, \bibinfo {author}
  {\bibfnamefont {A.}~\bibnamefont {Kreyssig}}, \bibinfo {author}
  {\bibfnamefont {A.}~\bibnamefont {Goldman}}, \bibinfo {author} {\bibfnamefont
  {R.}~\bibnamefont {McQueeney}}, \bibinfo {author} {\bibfnamefont
  {M.}~\bibnamefont {Torikachvili}}, \bibinfo {author} {\bibfnamefont
  {D.}~\bibnamefont {Argyriou}}, \bibinfo {author} {\bibfnamefont
  {G.}~\bibnamefont {Luke}}, \ and\ \bibinfo {author} {\bibfnamefont
  {W.}~\bibnamefont {Yu}},\ }\href {\doibase
  http://dx.doi.org/10.1016/j.physc.2009.03.033} {\bibfield  {journal}
  {\bibinfo  {journal} {Physica C: Superconductivity}\ }\textbf {\bibinfo
  {volume} {469}},\ \bibinfo {pages} {404 } (\bibinfo {year} {2009})},\
  \bibinfo {note} {superconductivity in Iron-Pnictides}\BibitemShut {NoStop}%
\bibitem [{\citenamefont {Kreyssig}\ \emph {et~al.}(2008)\citenamefont
  {Kreyssig}, \citenamefont {Green}, \citenamefont {Lee}, \citenamefont
  {Samolyuk}, \citenamefont {Zajdel}, \citenamefont {Lynn}, \citenamefont
  {Bud'ko}, \citenamefont {Torikachvili}, \citenamefont {Ni}, \citenamefont
  {Nandi}, \citenamefont {Le\~ao}, \citenamefont {Poulton}, \citenamefont
  {Argyriou}, \citenamefont {Harmon}, \citenamefont {McQueeney}, \citenamefont
  {Canfield},\ and\ \citenamefont {Goldman}}]{Kreyssig08}%
  \BibitemOpen
  \bibfield  {author} {\bibinfo {author} {\bibfnamefont {A.}~\bibnamefont
  {Kreyssig}}, \bibinfo {author} {\bibfnamefont {M.~A.}\ \bibnamefont {Green}},
  \bibinfo {author} {\bibfnamefont {Y.}~\bibnamefont {Lee}}, \bibinfo {author}
  {\bibfnamefont {G.~D.}\ \bibnamefont {Samolyuk}}, \bibinfo {author}
  {\bibfnamefont {P.}~\bibnamefont {Zajdel}}, \bibinfo {author} {\bibfnamefont
  {J.~W.}\ \bibnamefont {Lynn}}, \bibinfo {author} {\bibfnamefont {S.~L.}\
  \bibnamefont {Bud'ko}}, \bibinfo {author} {\bibfnamefont {M.~S.}\
  \bibnamefont {Torikachvili}}, \bibinfo {author} {\bibfnamefont
  {N.}~\bibnamefont {Ni}}, \bibinfo {author} {\bibfnamefont {S.}~\bibnamefont
  {Nandi}}, \bibinfo {author} {\bibfnamefont {J.~B.}\ \bibnamefont {Le\~ao}},
  \bibinfo {author} {\bibfnamefont {S.~J.}\ \bibnamefont {Poulton}}, \bibinfo
  {author} {\bibfnamefont {D.~N.}\ \bibnamefont {Argyriou}}, \bibinfo {author}
  {\bibfnamefont {B.~N.}\ \bibnamefont {Harmon}}, \bibinfo {author}
  {\bibfnamefont {R.~J.}\ \bibnamefont {McQueeney}}, \bibinfo {author}
  {\bibfnamefont {P.~C.}\ \bibnamefont {Canfield}}, \ and\ \bibinfo {author}
  {\bibfnamefont {A.~I.}\ \bibnamefont {Goldman}},\ }\href {\doibase
  10.1103/PhysRevB.78.184517} {\bibfield  {journal} {\bibinfo  {journal} {Phys.
  Rev. B}\ }\textbf {\bibinfo {volume} {78}},\ \bibinfo {pages} {184517}
  (\bibinfo {year} {2008})}\BibitemShut {NoStop}%
\bibitem [{\citenamefont {Ran}\ \emph {et~al.}(2011)\citenamefont {Ran},
  \citenamefont {Bud'ko}, \citenamefont {Pratt}, \citenamefont {Kreyssig},
  \citenamefont {Kim}, \citenamefont {Kramer}, \citenamefont {Ryan},
  \citenamefont {Rowan-Weetaluktuk}, \citenamefont {Furukawa}, \citenamefont
  {Roy}, \citenamefont {Goldman},\ and\ \citenamefont {Canfield}}]{Ran11}%
  \BibitemOpen
  \bibfield  {author} {\bibinfo {author} {\bibfnamefont {S.}~\bibnamefont
  {Ran}}, \bibinfo {author} {\bibfnamefont {S.~L.}\ \bibnamefont {Bud'ko}},
  \bibinfo {author} {\bibfnamefont {D.~K.}\ \bibnamefont {Pratt}}, \bibinfo
  {author} {\bibfnamefont {A.}~\bibnamefont {Kreyssig}}, \bibinfo {author}
  {\bibfnamefont {M.~G.}\ \bibnamefont {Kim}}, \bibinfo {author} {\bibfnamefont
  {M.~J.}\ \bibnamefont {Kramer}}, \bibinfo {author} {\bibfnamefont {D.~H.}\
  \bibnamefont {Ryan}}, \bibinfo {author} {\bibfnamefont {W.~N.}\ \bibnamefont
  {Rowan-Weetaluktuk}}, \bibinfo {author} {\bibfnamefont {Y.}~\bibnamefont
  {Furukawa}}, \bibinfo {author} {\bibfnamefont {B.}~\bibnamefont {Roy}},
  \bibinfo {author} {\bibfnamefont {A.~I.}\ \bibnamefont {Goldman}}, \ and\
  \bibinfo {author} {\bibfnamefont {P.~C.}\ \bibnamefont {Canfield}},\ }\href
  {\doibase 10.1103/PhysRevB.83.144517} {\bibfield  {journal} {\bibinfo
  {journal} {Phys. Rev. B}\ }\textbf {\bibinfo {volume} {83}},\ \bibinfo
  {pages} {144517} (\bibinfo {year} {2011})}\BibitemShut {NoStop}%
\bibitem [{\citenamefont {Ran}\ \emph {et~al.}(2012)\citenamefont {Ran},
  \citenamefont {Bud'ko}, \citenamefont {Straszheim}, \citenamefont {Soh},
  \citenamefont {Kim}, \citenamefont {Kreyssig}, \citenamefont {Goldman},\ and\
  \citenamefont {Canfield}}]{Ran12}%
  \BibitemOpen
  \bibfield  {author} {\bibinfo {author} {\bibfnamefont {S.}~\bibnamefont
  {Ran}}, \bibinfo {author} {\bibfnamefont {S.~L.}\ \bibnamefont {Bud'ko}},
  \bibinfo {author} {\bibfnamefont {W.~E.}\ \bibnamefont {Straszheim}},
  \bibinfo {author} {\bibfnamefont {J.}~\bibnamefont {Soh}}, \bibinfo {author}
  {\bibfnamefont {M.~G.}\ \bibnamefont {Kim}}, \bibinfo {author} {\bibfnamefont
  {A.}~\bibnamefont {Kreyssig}}, \bibinfo {author} {\bibfnamefont {A.~I.}\
  \bibnamefont {Goldman}}, \ and\ \bibinfo {author} {\bibfnamefont {P.~C.}\
  \bibnamefont {Canfield}},\ }\href {\doibase 10.1103/PhysRevB.85.224528}
  {\bibfield  {journal} {\bibinfo  {journal} {Phys. Rev. B}\ }\textbf {\bibinfo
  {volume} {85}},\ \bibinfo {pages} {224528} (\bibinfo {year}
  {2012})}\BibitemShut {NoStop}%
\bibitem [{\citenamefont {Ran}\ \emph {et~al.}(2014)\citenamefont {Ran},
  \citenamefont {Bud'ko}, \citenamefont {Straszheim},\ and\ \citenamefont
  {Canfield}}]{Ran14a}%
  \BibitemOpen
  \bibfield  {author} {\bibinfo {author} {\bibfnamefont {S.}~\bibnamefont
  {Ran}}, \bibinfo {author} {\bibfnamefont {S.~L.}\ \bibnamefont {Bud'ko}},
  \bibinfo {author} {\bibfnamefont {W.~E.}\ \bibnamefont {Straszheim}}, \ and\
  \bibinfo {author} {\bibfnamefont {P.~C.}\ \bibnamefont {Canfield}},\ }\href
  {\doibase 10.1103/PhysRevB.90.054501} {\bibfield  {journal} {\bibinfo
  {journal} {Phys. Rev. B}\ }\textbf {\bibinfo {volume} {90}},\ \bibinfo
  {pages} {054501} (\bibinfo {year} {2014})}\BibitemShut {NoStop}%
\bibitem [{\citenamefont {Ran}(2014)}]{Ran14b}%
  \BibitemOpen
  \bibfield  {author} {\bibinfo {author} {\bibfnamefont {S.}~\bibnamefont
  {Ran}},\ }\emph {\bibinfo {title} {Combined effects of post-growth thermal
  treatment and chemical substitution on physical properties of
  CaFe$_2$As$_2$}},\ \href@noop {} {Ph.D. thesis},\ \bibinfo  {school} {Iowa
  State University} (\bibinfo {year} {2014})\BibitemShut {NoStop}%
\bibitem [{\citenamefont {Iyo}\ \emph {et~al.}(2016)\citenamefont {Iyo},
  \citenamefont {Kawashima}, \citenamefont {Kinjo}, \citenamefont {Nishio},
  \citenamefont {Ishida}, \citenamefont {Fujihisa}, \citenamefont {Gotoh},
  \citenamefont {Kihou}, \citenamefont {Eisaki},\ and\ \citenamefont
  {Yoshida}}]{Iyo16}%
  \BibitemOpen
  \bibfield  {author} {\bibinfo {author} {\bibfnamefont {A.}~\bibnamefont
  {Iyo}}, \bibinfo {author} {\bibfnamefont {K.}~\bibnamefont {Kawashima}},
  \bibinfo {author} {\bibfnamefont {T.}~\bibnamefont {Kinjo}}, \bibinfo
  {author} {\bibfnamefont {T.}~\bibnamefont {Nishio}}, \bibinfo {author}
  {\bibfnamefont {S.}~\bibnamefont {Ishida}}, \bibinfo {author} {\bibfnamefont
  {H.}~\bibnamefont {Fujihisa}}, \bibinfo {author} {\bibfnamefont
  {Y.}~\bibnamefont {Gotoh}}, \bibinfo {author} {\bibfnamefont
  {K.}~\bibnamefont {Kihou}}, \bibinfo {author} {\bibfnamefont
  {H.}~\bibnamefont {Eisaki}}, \ and\ \bibinfo {author} {\bibfnamefont
  {Y.}~\bibnamefont {Yoshida}},\ }\href {\doibase 10.1021/jacs.5b12571}
  {\bibfield  {journal} {\bibinfo  {journal} {J. Am. Chem. Soc.}\ }\textbf
  {\bibinfo {volume} {138}},\ \bibinfo {pages} {3410} (\bibinfo {year}
  {2016})}\BibitemShut {NoStop}%
\bibitem [{\citenamefont {Wang}\ \emph {et~al.}(2013)\citenamefont {Wang},
  \citenamefont {Shangguan}, \citenamefont {He}, \citenamefont {Zhao},
  \citenamefont {Long}, \citenamefont {Wang},\ and\ \citenamefont
  {Wang}}]{Wang13}%
  \BibitemOpen
  \bibfield  {author} {\bibinfo {author} {\bibfnamefont {D.~M.}\ \bibnamefont
  {Wang}}, \bibinfo {author} {\bibfnamefont {X.~C.}\ \bibnamefont {Shangguan}},
  \bibinfo {author} {\bibfnamefont {J.~B.}\ \bibnamefont {He}}, \bibinfo
  {author} {\bibfnamefont {L.~X.}\ \bibnamefont {Zhao}}, \bibinfo {author}
  {\bibfnamefont {Y.~J.}\ \bibnamefont {Long}}, \bibinfo {author}
  {\bibfnamefont {P.~P.}\ \bibnamefont {Wang}}, \ and\ \bibinfo {author}
  {\bibfnamefont {L.}~\bibnamefont {Wang}},\ }\href {\doibase
  10.1007/s10948-013-2169-5} {\bibfield  {journal} {\bibinfo  {journal} {J.
  Supercond. Nov. Magn.}\ }\textbf {\bibinfo {volume} {26}},\ \bibinfo {pages}
  {2121} (\bibinfo {year} {2013})}\BibitemShut {NoStop}%
\bibitem [{\citenamefont {Kong}\ \emph {et~al.}(2015)\citenamefont {Kong},
  \citenamefont {Bud'ko},\ and\ \citenamefont {Canfield}}]{Kong15a}%
  \BibitemOpen
  \bibfield  {author} {\bibinfo {author} {\bibfnamefont {T.}~\bibnamefont
  {Kong}}, \bibinfo {author} {\bibfnamefont {S.~L.}\ \bibnamefont {Bud'ko}}, \
  and\ \bibinfo {author} {\bibfnamefont {P.~C.}\ \bibnamefont {Canfield}},\
  }\href {\doibase 10.1103/PhysRevB.91.020507} {\bibfield  {journal} {\bibinfo
  {journal} {Phys. Rev. B}\ }\textbf {\bibinfo {volume} {91}},\ \bibinfo
  {pages} {020507} (\bibinfo {year} {2015})}\BibitemShut {NoStop}%
\bibitem [{\citenamefont {Canfield}\ \emph {et~al.}(2016)\citenamefont
  {Canfield}, \citenamefont {Kong}, \citenamefont {Kaluarachchi},\ and\
  \citenamefont {Jo}}]{Canfield16}%
  \BibitemOpen
  \bibfield  {author} {\bibinfo {author} {\bibfnamefont {P.~C.}\ \bibnamefont
  {Canfield}}, \bibinfo {author} {\bibfnamefont {T.}~\bibnamefont {Kong}},
  \bibinfo {author} {\bibfnamefont {U.~S.}\ \bibnamefont {Kaluarachchi}}, \
  and\ \bibinfo {author} {\bibfnamefont {N.~H.}\ \bibnamefont {Jo}},\ }\href
  {\doibase 10.1080/14786435.2015.1122248} {\bibfield  {journal} {\bibinfo
  {journal} {Philos. Mag.}\ }\textbf {\bibinfo {volume} {96}},\ \bibinfo
  {pages} {84} (\bibinfo {year} {2016})}\BibitemShut {NoStop}%
\bibitem [{\citenamefont {Canfield}\ and\ \citenamefont
  {Fisk}(1992)}]{Canfield92}%
  \BibitemOpen
  \bibfield  {author} {\bibinfo {author} {\bibfnamefont {P.~C.}\ \bibnamefont
  {Canfield}}\ and\ \bibinfo {author} {\bibfnamefont {Z.}~\bibnamefont
  {Fisk}},\ }\href {\doibase 10.1080/13642819208215073} {\bibfield  {journal}
  {\bibinfo  {journal} {Phil. Mag. B}\ }\textbf {\bibinfo {volume} {65}},\
  \bibinfo {pages} {1117} (\bibinfo {year} {1992})}\BibitemShut {NoStop}%
\bibitem{Canfield10b} P. C. Canfield, in Book Series on Complex Metallic Alloys, Volume 2: Properties and Applications of Complex Intermetallics, Chapter 2, pp. 93-111 (World Scientiﬁc, Singapore, 2010)
  
\bibitem [{\citenamefont {Rotter}\ \emph
  {et~al.}(2008{\natexlab{b}})\citenamefont {Rotter}, \citenamefont {Pangerl},
  \citenamefont {Tegel},\ and\ \citenamefont {Johrendt}}]{Rotter08a}%
  \BibitemOpen
  \bibfield  {author} {\bibinfo {author} {\bibfnamefont {M.}~\bibnamefont
  {Rotter}}, \bibinfo {author} {\bibfnamefont {M.}~\bibnamefont {Pangerl}},
  \bibinfo {author} {\bibfnamefont {M.}~\bibnamefont {Tegel}}, \ and\ \bibinfo
  {author} {\bibfnamefont {D.}~\bibnamefont {Johrendt}},\ }\href {\doibase
  10.1002/anie.200803641} {\bibfield  {journal} {\bibinfo  {journal} {Angew.
  Chem. Int. Ed.}\ }\textbf {\bibinfo {volume} {47}},\ \bibinfo {pages} {7949}
  (\bibinfo {year} {2008}{\natexlab{b}})}\BibitemShut {NoStop}%
\bibitem [{\citenamefont {Jesche}\ \emph {et~al.}(2016)\citenamefont {Jesche},
  \citenamefont {Fix}, \citenamefont {Kreyssig}, \citenamefont {Meier},\ and\
  \citenamefont {Canfield}}]{Jesche16}%
  \BibitemOpen
  \bibfield  {author} {\bibinfo {author} {\bibfnamefont {A.}~\bibnamefont
  {Jesche}}, \bibinfo {author} {\bibfnamefont {M.}~\bibnamefont {Fix}},
  \bibinfo {author} {\bibfnamefont {A.}~\bibnamefont {Kreyssig}}, \bibinfo
  {author} {\bibfnamefont {W.~R.}\ \bibnamefont {Meier}}, \ and\ \bibinfo
  {author} {\bibfnamefont {P.~C.}\ \bibnamefont {Canfield}},\ }\href {\doibase
  10.1080/14786435.2016.1192725} {\bibfield  {journal} {\bibinfo  {journal}
  {Philos. Mag.}\ }\textbf {\bibinfo {volume} {96}},\ \bibinfo {pages} {2115}
  (\bibinfo {year} {2016})}\BibitemShut {NoStop}%
\bibitem [{\citenamefont {Kreyssig}\ \emph {et~al.}(2007)\citenamefont
  {Kreyssig}, \citenamefont {Chang}, \citenamefont {Janssen}, \citenamefont
  {Kim}, \citenamefont {Nandi}, \citenamefont {Yan}, \citenamefont {Tan},
  \citenamefont {McQueeney}, \citenamefont {Canfield},\ and\ \citenamefont
  {Goldman}}]{Kreyssig07}%
  \BibitemOpen
  \bibfield  {author} {\bibinfo {author} {\bibfnamefont {A.}~\bibnamefont
  {Kreyssig}}, \bibinfo {author} {\bibfnamefont {S.}~\bibnamefont {Chang}},
  \bibinfo {author} {\bibfnamefont {Y.}~\bibnamefont {Janssen}}, \bibinfo
  {author} {\bibfnamefont {J.~W.}\ \bibnamefont {Kim}}, \bibinfo {author}
  {\bibfnamefont {S.}~\bibnamefont {Nandi}}, \bibinfo {author} {\bibfnamefont
  {J.~Q.}\ \bibnamefont {Yan}}, \bibinfo {author} {\bibfnamefont
  {L.}~\bibnamefont {Tan}}, \bibinfo {author} {\bibfnamefont {R.~J.}\
  \bibnamefont {McQueeney}}, \bibinfo {author} {\bibfnamefont {P.~C.}\
  \bibnamefont {Canfield}}, \ and\ \bibinfo {author} {\bibfnamefont {A.~I.}\
  \bibnamefont {Goldman}},\ }\href {\doibase 10.1103/PhysRevB.76.054421}
  {\bibfield  {journal} {\bibinfo  {journal} {Phys. Rev. B}\ }\textbf {\bibinfo
  {volume} {76}},\ \bibinfo {pages} {054421} (\bibinfo {year}
  {2007})}\BibitemShut {NoStop}%
\bibitem [{\citenamefont {Mun}\ \emph {et~al.}(2010)\citenamefont {Mun},
  \citenamefont {Bud'ko}, \citenamefont {Torikachvili},\ and\ \citenamefont
  {Canfield}}]{Mun10b}%
  \BibitemOpen
  \bibfield  {author} {\bibinfo {author} {\bibfnamefont {E.}~\bibnamefont
  {Mun}}, \bibinfo {author} {\bibfnamefont {S.~L.}\ \bibnamefont {Bud'ko}},
  \bibinfo {author} {\bibfnamefont {M.~S.}\ \bibnamefont {Torikachvili}}, \
  and\ \bibinfo {author} {\bibfnamefont {P.~C.}\ \bibnamefont {Canfield}},\
  }\href {http://stacks.iop.org/0957-0233/21/i=5/a=055104} {\bibfield
  {journal} {\bibinfo  {journal} {Meas. Sci. Technol.}\ }\textbf {\bibinfo
  {volume} {21}},\ \bibinfo {pages} {055104} (\bibinfo {year}
  {2010})}\BibitemShut {NoStop}%
\bibitem [{\citenamefont {Prozorov}\ \emph {et~al.}(2009)\citenamefont
  {Prozorov}, \citenamefont {Tillman}, \citenamefont {Mun},\ and\ \citenamefont
  {Canfield}}]{Prozorov2009d}%
  \BibitemOpen
  \bibfield  {author} {\bibinfo {author} {\bibfnamefont {R.}~\bibnamefont
  {Prozorov}}, \bibinfo {author} {\bibfnamefont {M.~E.}\ \bibnamefont
  {Tillman}}, \bibinfo {author} {\bibfnamefont {E.~D.}\ \bibnamefont {Mun}}, \
  and\ \bibinfo {author} {\bibfnamefont {P.~C.}\ \bibnamefont {Canfield}},\
  }\href {http://stacks.iop.org/1367-2630/11/i=3/a=035004} {\bibfield
  {journal} {\bibinfo  {journal} {New J. Phys.}\ }\textbf {\bibinfo {volume}
  {11}},\ \bibinfo {pages} {035004} (\bibinfo {year} {2009})}\BibitemShut
  {NoStop}%
\bibitem [{\citenamefont {Prozorov}\ \emph {et~al.}(2008)\citenamefont
  {Prozorov}, \citenamefont {Ni}, \citenamefont {Tanatar}, \citenamefont
  {Kogan}, \citenamefont {Gordon}, \citenamefont {Martin}, \citenamefont
  {Blomberg}, \citenamefont {Prommapan}, \citenamefont {Yan}, \citenamefont
  {Bud'ko},\ and\ \citenamefont {Canfield}}]{Prozorov2008a}%
  \BibitemOpen
  \bibfield  {author} {\bibinfo {author} {\bibfnamefont {R.}~\bibnamefont
  {Prozorov}}, \bibinfo {author} {\bibfnamefont {N.}~\bibnamefont {Ni}},
  \bibinfo {author} {\bibfnamefont {M.~A.}\ \bibnamefont {Tanatar}}, \bibinfo
  {author} {\bibfnamefont {V.~G.}\ \bibnamefont {Kogan}}, \bibinfo {author}
  {\bibfnamefont {R.~T.}\ \bibnamefont {Gordon}}, \bibinfo {author}
  {\bibfnamefont {C.}~\bibnamefont {Martin}}, \bibinfo {author} {\bibfnamefont
  {E.~C.}\ \bibnamefont {Blomberg}}, \bibinfo {author} {\bibfnamefont
  {P.}~\bibnamefont {Prommapan}}, \bibinfo {author} {\bibfnamefont {J.~Q.}\
  \bibnamefont {Yan}}, \bibinfo {author} {\bibfnamefont {S.~L.}\ \bibnamefont
  {Bud'ko}}, \ and\ \bibinfo {author} {\bibfnamefont {P.~C.}\ \bibnamefont
  {Canfield}},\ }\href {\doibase 10.1103/PhysRevB.78.224506} {\bibfield
  {journal} {\bibinfo  {journal} {Phys. Rev. B}\ }\textbf {\bibinfo {volume}
  {78}},\ \bibinfo {pages} {224506} (\bibinfo {year} {2008})}\BibitemShut
  {NoStop}%
\bibitem [{\citenamefont {Prozorov}\ \emph {et~al.}(2010)\citenamefont
  {Prozorov}, \citenamefont {Tanatar}, \citenamefont {Roy}, \citenamefont {Ni},
  \citenamefont {Bud'ko}, \citenamefont {Canfield}, \citenamefont {Hua},
  \citenamefont {Welp},\ and\ \citenamefont {Kwok}}]{Prozorov2010a}%
  \BibitemOpen
  \bibfield  {author} {\bibinfo {author} {\bibfnamefont {R.}~\bibnamefont
  {Prozorov}}, \bibinfo {author} {\bibfnamefont {M.~A.}\ \bibnamefont
  {Tanatar}}, \bibinfo {author} {\bibfnamefont {B.}~\bibnamefont {Roy}},
  \bibinfo {author} {\bibfnamefont {N.}~\bibnamefont {Ni}}, \bibinfo {author}
  {\bibfnamefont {S.~L.}\ \bibnamefont {Bud'ko}}, \bibinfo {author}
  {\bibfnamefont {P.~C.}\ \bibnamefont {Canfield}}, \bibinfo {author}
  {\bibfnamefont {J.}~\bibnamefont {Hua}}, \bibinfo {author} {\bibfnamefont
  {U.}~\bibnamefont {Welp}}, \ and\ \bibinfo {author} {\bibfnamefont {W.~K.}\
  \bibnamefont {Kwok}},\ }\href {\doibase 10.1103/PhysRevB.81.094509}
  {\bibfield  {journal} {\bibinfo  {journal} {Phys. Rev. B}\ }\textbf {\bibinfo
  {volume} {81}},\ \bibinfo {pages} {094509} (\bibinfo {year}
  {2010})}\BibitemShut {NoStop}%
\bibitem [{\citenamefont {Eiling}\ and\ \citenamefont
  {Schilling}(1981)}]{Schilling81}%
  \BibitemOpen
  \bibfield  {author} {\bibinfo {author} {\bibfnamefont {A.}~\bibnamefont
  {Eiling}}\ and\ \bibinfo {author} {\bibfnamefont {J.~S.}\ \bibnamefont
  {Schilling}},\ }\href {http://stacks.iop.org/0305-4608/11/i=3/a=010}
  {\bibfield  {journal} {\bibinfo  {journal} {J. Phys. F: Met. Phys.}\ }\textbf
  {\bibinfo {volume} {11}},\ \bibinfo {pages} {623} (\bibinfo {year}
  {1981})}\BibitemShut {NoStop}%
\bibitem [{\citenamefont {Alireza}\ \emph {et~al.}(2007)\citenamefont
  {Alireza}, \citenamefont {Barakat}, \citenamefont {Cumberlidge},
  \citenamefont {Lonzarich}, \citenamefont {Nakamura},\ and\ \citenamefont
  {Maeno}}]{Alireza07}%
  \BibitemOpen
  \bibfield  {author} {\bibinfo {author} {\bibfnamefont {P.~L.}\ \bibnamefont
  {Alireza}}, \bibinfo {author} {\bibfnamefont {S.}~\bibnamefont {Barakat}},
  \bibinfo {author} {\bibfnamefont {A.-M.}\ \bibnamefont {Cumberlidge}},
  \bibinfo {author} {\bibfnamefont {G.}~\bibnamefont {Lonzarich}}, \bibinfo
  {author} {\bibfnamefont {F.}~\bibnamefont {Nakamura}}, \ and\ \bibinfo
  {author} {\bibfnamefont {Y.}~\bibnamefont {Maeno}},\ }\href {\doibase
  10.1143/JPSJS.76SA.216} {\bibfield  {journal} {\bibinfo  {journal} {J. Phys.
  Soc. Jpn.}\ }\textbf {\bibinfo {volume} {76}},\ \bibinfo {pages} {216}
  (\bibinfo {year} {2007})}\BibitemShut {NoStop}%
\bibitem [{\citenamefont {Tanatar}\ \emph
  {et~al.}(2009{\natexlab{a}})\citenamefont {Tanatar}, \citenamefont {Ni},
  \citenamefont {Martin}, \citenamefont {Gordon}, \citenamefont {Kim},
  \citenamefont {Kogan}, \citenamefont {Samolyuk}, \citenamefont {Bud'ko},
  \citenamefont {Canfield},\ and\ \citenamefont {Prozorov}}]{Tanatar09}%
  \BibitemOpen
  \bibfield  {author} {\bibinfo {author} {\bibfnamefont {M.~A.}\ \bibnamefont
  {Tanatar}}, \bibinfo {author} {\bibfnamefont {N.}~\bibnamefont {Ni}},
  \bibinfo {author} {\bibfnamefont {C.}~\bibnamefont {Martin}}, \bibinfo
  {author} {\bibfnamefont {R.~T.}\ \bibnamefont {Gordon}}, \bibinfo {author}
  {\bibfnamefont {H.}~\bibnamefont {Kim}}, \bibinfo {author} {\bibfnamefont
  {V.~G.}\ \bibnamefont {Kogan}}, \bibinfo {author} {\bibfnamefont {G.~D.}\
  \bibnamefont {Samolyuk}}, \bibinfo {author} {\bibfnamefont {S.~L.}\
  \bibnamefont {Bud'ko}}, \bibinfo {author} {\bibfnamefont {P.~C.}\
  \bibnamefont {Canfield}}, \ and\ \bibinfo {author} {\bibfnamefont
  {R.}~\bibnamefont {Prozorov}},\ }\href {\doibase 10.1103/PhysRevB.79.094507}
  {\bibfield  {journal} {\bibinfo  {journal} {Phys. Rev. B}\ }\textbf {\bibinfo
  {volume} {79}},\ \bibinfo {pages} {094507} (\bibinfo {year}
  {2009}{\natexlab{a}})}\BibitemShut {NoStop}%
\bibitem [{\citenamefont {Tanatar}\ \emph {et~al.}(2010)\citenamefont
  {Tanatar}, \citenamefont {Ni}, \citenamefont {Bud'ko}, \citenamefont
  {Canfield},\ and\ \citenamefont {Prozorov}}]{Tanatar10}%
  \BibitemOpen
  \bibfield  {author} {\bibinfo {author} {\bibfnamefont {M.~A.}\ \bibnamefont
  {Tanatar}}, \bibinfo {author} {\bibfnamefont {N.}~\bibnamefont {Ni}},
  \bibinfo {author} {\bibfnamefont {S.~L.}\ \bibnamefont {Bud'ko}}, \bibinfo
  {author} {\bibfnamefont {P.~C.}\ \bibnamefont {Canfield}}, \ and\ \bibinfo
  {author} {\bibfnamefont {R.}~\bibnamefont {Prozorov}},\ }\href
  {http://stacks.iop.org/0953-2048/23/i=5/a=054002} {\bibfield  {journal}
  {\bibinfo  {journal} {Supercond. Sci. Technol.}\ }\textbf {\bibinfo {volume}
  {23}},\ \bibinfo {pages} {054002} (\bibinfo {year} {2010})}\BibitemShut
  {NoStop}%
\bibitem [{\citenamefont {Tanatar}\ \emph {et~al.}(2013)\citenamefont
  {Tanatar}, \citenamefont {Prozorov}, \citenamefont {Ni}, \citenamefont
  {Bud'ko},\ and\ \citenamefont {Canfield}}]{Tanatar}%
  \BibitemOpen
  \bibfield  {author} {\bibinfo {author} {\bibfnamefont {M.}~\bibnamefont
  {Tanatar}}, \bibinfo {author} {\bibfnamefont {R.}~\bibnamefont {Prozorov}},
  \bibinfo {author} {\bibfnamefont {N.}~\bibnamefont {Ni}}, \bibinfo {author}
  {\bibfnamefont {S.}~\bibnamefont {Bud'ko}}, \ and\ \bibinfo {author}
  {\bibfnamefont {P.}~\bibnamefont {Canfield}},\ }\href@noop {} {\enquote
  {\bibinfo {title} {Low resistivity contact to iron-pnictide
  superconductors},}\ } (\bibinfo {year} {2013}),\ \bibinfo {note} {uS Patent
  8,450,246}\BibitemShut {NoStop}%
\bibitem [{\citenamefont {Tanatar}\ \emph
  {et~al.}(2009{\natexlab{b}})\citenamefont {Tanatar}, \citenamefont {Ni},
  \citenamefont {Samolyuk}, \citenamefont {Bud'ko}, \citenamefont {Canfield},\
  and\ \citenamefont {Prozorov}}]{Tanatar09b}%
  \BibitemOpen
  \bibfield  {author} {\bibinfo {author} {\bibfnamefont {M.~A.}\ \bibnamefont
  {Tanatar}}, \bibinfo {author} {\bibfnamefont {N.}~\bibnamefont {Ni}},
  \bibinfo {author} {\bibfnamefont {G.~D.}\ \bibnamefont {Samolyuk}}, \bibinfo
  {author} {\bibfnamefont {S.~L.}\ \bibnamefont {Bud'ko}}, \bibinfo {author}
  {\bibfnamefont {P.~C.}\ \bibnamefont {Canfield}}, \ and\ \bibinfo {author}
  {\bibfnamefont {R.}~\bibnamefont {Prozorov}},\ }\href {\doibase
  10.1103/PhysRevB.79.134528} {\bibfield  {journal} {\bibinfo  {journal} {Phys.
  Rev. B}\ }\textbf {\bibinfo {volume} {79}},\ \bibinfo {pages} {134528}
  (\bibinfo {year} {2009}{\natexlab{b}})}\BibitemShut {NoStop}%
\bibitem [{\citenamefont {Chu}\ \emph {et~al.}(2012)\citenamefont {Chu},
  \citenamefont {Kuo}, \citenamefont {Analytis},\ and\ \citenamefont
  {Fisher}}]{Chu12}%
  \BibitemOpen
  \bibfield  {author} {\bibinfo {author} {\bibfnamefont {J.-H.}\ \bibnamefont
  {Chu}}, \bibinfo {author} {\bibfnamefont {H.-H.}\ \bibnamefont {Kuo}},
  \bibinfo {author} {\bibfnamefont {J.~G.}\ \bibnamefont {Analytis}}, \ and\
  \bibinfo {author} {\bibfnamefont {I.~R.}\ \bibnamefont {Fisher}},\ }\href
  {\doibase 10.1126/science.1221713} {\bibfield  {journal} {\bibinfo  {journal}
  {Science}\ }\textbf {\bibinfo {volume} {337}},\ \bibinfo {pages} {710}
  (\bibinfo {year} {2012})}\BibitemShut {NoStop}%
\bibitem [{\citenamefont {Kuo}\ \emph {et~al.}(2013)\citenamefont {Kuo},
  \citenamefont {Shapiro}, \citenamefont {Riggs},\ and\ \citenamefont
  {Fisher}}]{Kuo13}%
  \BibitemOpen
  \bibfield  {author} {\bibinfo {author} {\bibfnamefont {H.-H.}\ \bibnamefont
  {Kuo}}, \bibinfo {author} {\bibfnamefont {M.~C.}\ \bibnamefont {Shapiro}},
  \bibinfo {author} {\bibfnamefont {S.~C.}\ \bibnamefont {Riggs}}, \ and\
  \bibinfo {author} {\bibfnamefont {I.~R.}\ \bibnamefont {Fisher}},\ }\href
  {\doibase 10.1103/PhysRevB.88.085113} {\bibfield  {journal} {\bibinfo
  {journal} {Phys. Rev. B}\ }\textbf {\bibinfo {volume} {88}},\ \bibinfo
  {pages} {085113} (\bibinfo {year} {2013})}\BibitemShut {NoStop}%
\bibitem [{\citenamefont {Mou}\ \emph {et~al.}(2016)\citenamefont {Mou},
  \citenamefont {Kong}, \citenamefont {Meier}, \citenamefont {Lochner},
  \citenamefont {Wang}, \citenamefont {Lin}, \citenamefont {Wu}, \citenamefont
  {Bud'ko}, \citenamefont {Eremin}, \citenamefont {Johnson}, \citenamefont
  {Canfield},\ and\ \citenamefont {Kaminski}}]{Mou16}%
  \BibitemOpen
  \bibfield  {author} {\bibinfo {author} {\bibfnamefont {D.}~\bibnamefont
  {Mou}}, \bibinfo {author} {\bibfnamefont {T.}~\bibnamefont {Kong}}, \bibinfo
  {author} {\bibfnamefont {W.~R.}\ \bibnamefont {Meier}}, \bibinfo {author}
  {\bibfnamefont {F.}~\bibnamefont {Lochner}}, \bibinfo {author} {\bibfnamefont
  {L.-L.}\ \bibnamefont {Wang}}, \bibinfo {author} {\bibfnamefont
  {Q.}~\bibnamefont {Lin}}, \bibinfo {author} {\bibfnamefont {Y.}~\bibnamefont
  {Wu}}, \bibinfo {author} {\bibfnamefont {S.}~\bibnamefont {Bud'ko}}, \bibinfo
  {author} {\bibfnamefont {I.}~\bibnamefont {Eremin}}, \bibinfo {author}
  {\bibfnamefont {D.}~\bibnamefont {Johnson}}, \bibinfo {author} {\bibfnamefont
  {P.~C.}\ \bibnamefont {Canfield}}, \ and\ \bibinfo {author} {\bibfnamefont
  {A.}~\bibnamefont {Kaminski}},\ }\href@noop {} {\bibfield  {journal}
  {\bibinfo  {journal} {arXiv:1606.05643}\ } (\bibinfo {year}
  {2016})}\BibitemShut {NoStop}%
\bibitem [{\citenamefont {Cho}\ \emph {et~al.}(2016)\citenamefont {Cho},
  \citenamefont {Fente}, \citenamefont {Teknowijoyo}, \citenamefont {Tanatar},
  \citenamefont {Kong}, \citenamefont {Meier}, \citenamefont {Kaluarachchi},
  \citenamefont {Guillam{\'o}n}, \citenamefont {Suderow}, \citenamefont
  {Bud'ko} \emph {et~al.}}]{Cho16}%
  \BibitemOpen
  \bibfield  {author} {\bibinfo {author} {\bibfnamefont {K.}~\bibnamefont
  {Cho}}, \bibinfo {author} {\bibfnamefont {A.}~\bibnamefont {Fente}}, \bibinfo
  {author} {\bibfnamefont {S.}~\bibnamefont {Teknowijoyo}}, \bibinfo {author}
  {\bibfnamefont {M.}~\bibnamefont {Tanatar}}, \bibinfo {author} {\bibfnamefont
  {T.}~\bibnamefont {Kong}}, \bibinfo {author} {\bibfnamefont {W.}~\bibnamefont
  {Meier}}, \bibinfo {author} {\bibfnamefont {U.}~\bibnamefont {Kaluarachchi}},
  \bibinfo {author} {\bibfnamefont {I.}~\bibnamefont {Guillam{\'o}n}}, \bibinfo
  {author} {\bibfnamefont {H.}~\bibnamefont {Suderow}}, \bibinfo {author}
  {\bibfnamefont {S.}~\bibnamefont {Bud'ko}},  \emph {et~al.},\ }\href@noop {}
  {\bibfield  {journal} {\bibinfo  {journal} {arXiv:1606.06245}\ } (\bibinfo
  {year} {2016})}\BibitemShut {NoStop}%
\bibitem [{\citenamefont {Ziman}(1972)}]{Ziman}%
  \BibitemOpen
  \bibfield  {author} {\bibinfo {author} {\bibfnamefont {J.~M.}\ \bibnamefont
  {Ziman}},\ }\href@noop {} {\emph {\bibinfo {title} {Principles of the Theory
  of Solids}}}\ (\bibinfo  {publisher} {Cambridge University Press},\ \bibinfo
  {year} {1972})\BibitemShut {NoStop}%
\bibitem [{\citenamefont {Ohgushi}\ and\ \citenamefont
  {Kiuchi}(2012)}]{Ohgushi12}%
  \BibitemOpen
  \bibfield  {author} {\bibinfo {author} {\bibfnamefont {K.}~\bibnamefont
  {Ohgushi}}\ and\ \bibinfo {author} {\bibfnamefont {Y.}~\bibnamefont
  {Kiuchi}},\ }\href {\doibase 10.1103/PhysRevB.85.064522} {\bibfield
  {journal} {\bibinfo  {journal} {Phys. Rev. B}\ }\textbf {\bibinfo {volume}
  {85}},\ \bibinfo {pages} {064522} (\bibinfo {year} {2012})}\BibitemShut
  {NoStop}%
\bibitem [{\citenamefont {Hodovanets}\ \emph {et~al.}(2014)\citenamefont
  {Hodovanets}, \citenamefont {Liu}, \citenamefont {Jesche}, \citenamefont
  {Ran}, \citenamefont {Mun}, \citenamefont {Lograsso}, \citenamefont
  {Bud'ko},\ and\ \citenamefont {Canfield}}]{Halyna14a}%
  \BibitemOpen
  \bibfield  {author} {\bibinfo {author} {\bibfnamefont {H.}~\bibnamefont
  {Hodovanets}}, \bibinfo {author} {\bibfnamefont {Y.}~\bibnamefont {Liu}},
  \bibinfo {author} {\bibfnamefont {A.}~\bibnamefont {Jesche}}, \bibinfo
  {author} {\bibfnamefont {S.}~\bibnamefont {Ran}}, \bibinfo {author}
  {\bibfnamefont {E.~D.}\ \bibnamefont {Mun}}, \bibinfo {author} {\bibfnamefont
  {T.~A.}\ \bibnamefont {Lograsso}}, \bibinfo {author} {\bibfnamefont {S.~L.}\
  \bibnamefont {Bud'ko}}, \ and\ \bibinfo {author} {\bibfnamefont {P.~C.}\
  \bibnamefont {Canfield}},\ }\href {\doibase 10.1103/PhysRevB.89.224517}
  {\bibfield  {journal} {\bibinfo  {journal} {Phys. Rev. B}\ }\textbf {\bibinfo
  {volume} {89}},\ \bibinfo {pages} {224517} (\bibinfo {year}
  {2014})}\BibitemShut {NoStop}%
\bibitem [{\citenamefont {Bean}(1964)}]{Bean64}%
  \BibitemOpen
  \bibfield  {author} {\bibinfo {author} {\bibfnamefont {C.~P.}\ \bibnamefont
  {Bean}},\ }\href {\doibase 10.1103/RevModPhys.36.31} {\bibfield  {journal}
  {\bibinfo  {journal} {Rev. Mod. Phys.}\ }\textbf {\bibinfo {volume} {36}},\
  \bibinfo {pages} {31} (\bibinfo {year} {1964})}\BibitemShut {NoStop}%
\bibitem [{\citenamefont {Welp}\ \emph {et~al.}(1989)\citenamefont {Welp},
  \citenamefont {Kwok}, \citenamefont {Crabtree}, \citenamefont {Vandervoort},\
  and\ \citenamefont {Liu}}]{Welp89}%
  \BibitemOpen
  \bibfield  {author} {\bibinfo {author} {\bibfnamefont {U.}~\bibnamefont
  {Welp}}, \bibinfo {author} {\bibfnamefont {W.~K.}\ \bibnamefont {Kwok}},
  \bibinfo {author} {\bibfnamefont {G.~W.}\ \bibnamefont {Crabtree}}, \bibinfo
  {author} {\bibfnamefont {K.~G.}\ \bibnamefont {Vandervoort}}, \ and\ \bibinfo
  {author} {\bibfnamefont {J.~Z.}\ \bibnamefont {Liu}},\ }\href {\doibase
  10.1103/PhysRevLett.62.1908} {\bibfield  {journal} {\bibinfo  {journal}
  {Phys. Rev. Lett.}\ }\textbf {\bibinfo {volume} {62}},\ \bibinfo {pages}
  {1908} (\bibinfo {year} {1989})}\BibitemShut {NoStop}%
\bibitem [{\citenamefont {Colombier}\ \emph {et~al.}(2010)\citenamefont
  {Colombier}, \citenamefont {Torikachvili}, \citenamefont {Ni}, \citenamefont
  {Thaler}, \citenamefont {Bud'ko},\ and\ \citenamefont
  {Canfield}}]{Colombier10}%
  \BibitemOpen
  \bibfield  {author} {\bibinfo {author} {\bibfnamefont {E.}~\bibnamefont
  {Colombier}}, \bibinfo {author} {\bibfnamefont {M.~S.}\ \bibnamefont
  {Torikachvili}}, \bibinfo {author} {\bibfnamefont {N.}~\bibnamefont {Ni}},
  \bibinfo {author} {\bibfnamefont {A.}~\bibnamefont {Thaler}}, \bibinfo
  {author} {\bibfnamefont {S.~L.}\ \bibnamefont {Bud'ko}}, \ and\ \bibinfo
  {author} {\bibfnamefont {P.~C.}\ \bibnamefont {Canfield}},\ }\href
  {http://stacks.iop.org/0953-2048/23/i=5/a=054003} {\bibfield  {journal}
  {\bibinfo  {journal} {Supercond. Sci. Technol.}\ }\textbf {\bibinfo {volume}
  {23}},\ \bibinfo {pages} {054003} (\bibinfo {year} {2010})}\BibitemShut
  {NoStop}%
\bibitem [{\citenamefont {Bud'ko}\ \emph {et~al.}(2013)\citenamefont {Bud'ko},
  \citenamefont {Sturza}, \citenamefont {Chung}, \citenamefont {Kanatzidis},\
  and\ \citenamefont {Canfield}}]{Sergey13}%
  \BibitemOpen
  \bibfield  {author} {\bibinfo {author} {\bibfnamefont {S.~L.}\ \bibnamefont
  {Bud'ko}}, \bibinfo {author} {\bibfnamefont {M.}~\bibnamefont {Sturza}},
  \bibinfo {author} {\bibfnamefont {D.~Y.}\ \bibnamefont {Chung}}, \bibinfo
  {author} {\bibfnamefont {M.~G.}\ \bibnamefont {Kanatzidis}}, \ and\ \bibinfo
  {author} {\bibfnamefont {P.~C.}\ \bibnamefont {Canfield}},\ }\href {\doibase
  10.1103/PhysRevB.87.100509} {\bibfield  {journal} {\bibinfo  {journal} {Phys.
  Rev. B}\ }\textbf {\bibinfo {volume} {87}},\ \bibinfo {pages} {100509}
  (\bibinfo {year} {2013})}\BibitemShut {NoStop}%
\bibitem [{\citenamefont {Gurevich}(2011)}]{Gurevich11}%
  \BibitemOpen
  \bibfield  {author} {\bibinfo {author} {\bibfnamefont {A.}~\bibnamefont
  {Gurevich}},\ }\href {http://stacks.iop.org/0034-4885/74/i=12/a=124501}
  {\bibfield  {journal} {\bibinfo  {journal} {Rep. Prog. Phys.}\ }\textbf
  {\bibinfo {volume} {74}},\ \bibinfo {pages} {124501} (\bibinfo {year}
  {2011})}\BibitemShut {NoStop}%
\bibitem [{\citenamefont {Altarawneh}\ \emph {et~al.}(2008)\citenamefont
  {Altarawneh}, \citenamefont {Collar}, \citenamefont {Mielke}, \citenamefont
  {Ni}, \citenamefont {Bud'ko},\ and\ \citenamefont {Canfield}}]{Altarawneh08}%
  \BibitemOpen
  \bibfield  {author} {\bibinfo {author} {\bibfnamefont {M.~M.}\ \bibnamefont
  {Altarawneh}}, \bibinfo {author} {\bibfnamefont {K.}~\bibnamefont {Collar}},
  \bibinfo {author} {\bibfnamefont {C.~H.}\ \bibnamefont {Mielke}}, \bibinfo
  {author} {\bibfnamefont {N.}~\bibnamefont {Ni}}, \bibinfo {author}
  {\bibfnamefont {S.~L.}\ \bibnamefont {Bud'ko}}, \ and\ \bibinfo {author}
  {\bibfnamefont {P.~C.}\ \bibnamefont {Canfield}},\ }\href {\doibase
  10.1103/PhysRevB.78.220505} {\bibfield  {journal} {\bibinfo  {journal} {Phys.
  Rev. B}\ }\textbf {\bibinfo {volume} {78}},\ \bibinfo {pages} {220505}
  (\bibinfo {year} {2008})}\BibitemShut {NoStop}%
\bibitem [{\citenamefont {Rutgers}(1934)}]{Rutgers34}%
  \BibitemOpen
  \bibfield  {author} {\bibinfo {author} {\bibfnamefont {A.}~\bibnamefont
  {Rutgers}},\ }\href {\doibase
  http://dx.doi.org/10.1016/S0031-8914(34)80300-X} {\bibfield  {journal}
  {\bibinfo  {journal} {Physica}\ }\textbf {\bibinfo {volume} {1}},\ \bibinfo
  {pages} {1055 } (\bibinfo {year} {1934})}\BibitemShut {NoStop}%
\bibitem [{\citenamefont {Hirschfeld}\ \emph {et~al.}(2011)\citenamefont
  {Hirschfeld}, \citenamefont {Korshunov},\ and\ \citenamefont
  {Mazin}}]{Hirschfeld11}%
  \BibitemOpen
  \bibfield  {author} {\bibinfo {author} {\bibfnamefont {P.~J.}\ \bibnamefont
  {Hirschfeld}}, \bibinfo {author} {\bibfnamefont {M.~M.}\ \bibnamefont
  {Korshunov}}, \ and\ \bibinfo {author} {\bibfnamefont {I.~I.}\ \bibnamefont
  {Mazin}},\ }\href {http://stacks.iop.org/0034-4885/74/i=12/a=124508}
  {\bibfield  {journal} {\bibinfo  {journal} {Rep. Prog. Phys.}\ }\textbf
  {\bibinfo {volume} {74}},\ \bibinfo {pages} {124508} (\bibinfo {year}
  {2011})}\BibitemShut {NoStop}%
\bibitem [{\citenamefont {Tarantini}\ \emph {et~al.}(2011)\citenamefont
  {Tarantini}, \citenamefont {Gurevich}, \citenamefont {Jaroszynski},
  \citenamefont {Balakirev}, \citenamefont {Bellingeri}, \citenamefont
  {Pallecchi}, \citenamefont {Ferdeghini}, \citenamefont {Shen}, \citenamefont
  {Wen},\ and\ \citenamefont {Larbalestier}}]{Tarantini11}%
  \BibitemOpen
  \bibfield  {author} {\bibinfo {author} {\bibfnamefont {C.}~\bibnamefont
  {Tarantini}}, \bibinfo {author} {\bibfnamefont {A.}~\bibnamefont {Gurevich}},
  \bibinfo {author} {\bibfnamefont {J.}~\bibnamefont {Jaroszynski}}, \bibinfo
  {author} {\bibfnamefont {F.}~\bibnamefont {Balakirev}}, \bibinfo {author}
  {\bibfnamefont {E.}~\bibnamefont {Bellingeri}}, \bibinfo {author}
  {\bibfnamefont {I.}~\bibnamefont {Pallecchi}}, \bibinfo {author}
  {\bibfnamefont {C.}~\bibnamefont {Ferdeghini}}, \bibinfo {author}
  {\bibfnamefont {B.}~\bibnamefont {Shen}}, \bibinfo {author} {\bibfnamefont
  {H.~H.}\ \bibnamefont {Wen}}, \ and\ \bibinfo {author} {\bibfnamefont
  {D.~C.}\ \bibnamefont {Larbalestier}},\ }\href {\doibase
  10.1103/PhysRevB.84.184522} {\bibfield  {journal} {\bibinfo  {journal} {Phys.
  Rev. B}\ }\textbf {\bibinfo {volume} {84}},\ \bibinfo {pages} {184522}
  (\bibinfo {year} {2011})}\BibitemShut {NoStop}%
\bibitem [{\citenamefont {Cho}\ \emph {et~al.}(2011)\citenamefont {Cho},
  \citenamefont {Kim}, \citenamefont {Tanatar}, \citenamefont {Song},
  \citenamefont {Kwon}, \citenamefont {Coniglio}, \citenamefont {Agosta},
  \citenamefont {Gurevich},\ and\ \citenamefont {Prozorov}}]{Cho11}%
  \BibitemOpen
  \bibfield  {author} {\bibinfo {author} {\bibfnamefont {K.}~\bibnamefont
  {Cho}}, \bibinfo {author} {\bibfnamefont {H.}~\bibnamefont {Kim}}, \bibinfo
  {author} {\bibfnamefont {M.~A.}\ \bibnamefont {Tanatar}}, \bibinfo {author}
  {\bibfnamefont {Y.~J.}\ \bibnamefont {Song}}, \bibinfo {author}
  {\bibfnamefont {Y.~S.}\ \bibnamefont {Kwon}}, \bibinfo {author}
  {\bibfnamefont {W.~A.}\ \bibnamefont {Coniglio}}, \bibinfo {author}
  {\bibfnamefont {C.~C.}\ \bibnamefont {Agosta}}, \bibinfo {author}
  {\bibfnamefont {A.}~\bibnamefont {Gurevich}}, \ and\ \bibinfo {author}
  {\bibfnamefont {R.}~\bibnamefont {Prozorov}},\ }\href {\doibase
  10.1103/PhysRevB.83.060502} {\bibfield  {journal} {\bibinfo  {journal} {Phys.
  Rev. B}\ }\textbf {\bibinfo {volume} {83}},\ \bibinfo {pages} {060502}
  (\bibinfo {year} {2011})}\BibitemShut {NoStop}%
\bibitem [{\citenamefont {Blatter}\ \emph {et~al.}(1994)\citenamefont
  {Blatter}, \citenamefont {Feigel'man}, \citenamefont {Geshkenbein},
  \citenamefont {Larkin},\ and\ \citenamefont {Vinokur}}]{Blatter94}%
  \BibitemOpen
  \bibfield  {author} {\bibinfo {author} {\bibfnamefont {G.}~\bibnamefont
  {Blatter}}, \bibinfo {author} {\bibfnamefont {M.~V.}\ \bibnamefont
  {Feigel'man}}, \bibinfo {author} {\bibfnamefont {V.~B.}\ \bibnamefont
  {Geshkenbein}}, \bibinfo {author} {\bibfnamefont {A.~I.}\ \bibnamefont
  {Larkin}}, \ and\ \bibinfo {author} {\bibfnamefont {V.~M.}\ \bibnamefont
  {Vinokur}},\ }\href {\doibase 10.1103/RevModPhys.66.1125} {\bibfield
  {journal} {\bibinfo  {journal} {Rev. Mod. Phys.}\ }\textbf {\bibinfo {volume}
  {66}},\ \bibinfo {pages} {1125} (\bibinfo {year} {1994})}\BibitemShut
  {NoStop}%
\bibitem [{\citenamefont {Putti}\ \emph {et~al.}(2010)\citenamefont {Putti},
  \citenamefont {Pallecchi}, \citenamefont {Bellingeri}, \citenamefont
  {Cimberle}, \citenamefont {Tropeano}, \citenamefont {Ferdeghini},
  \citenamefont {Palenzona}, \citenamefont {Tarantini}, \citenamefont
  {Yamamoto}, \citenamefont {Jiang}, \citenamefont {Jaroszynski}, \citenamefont
  {Kametani}, \citenamefont {Abraimov}, \citenamefont {Polyanskii},
  \citenamefont {Weiss}, \citenamefont {Hellstrom}, \citenamefont {Gurevich},
  \citenamefont {Larbalestier}, \citenamefont {Jin}, \citenamefont {Sales},
  \citenamefont {Sefat}, \citenamefont {McGuire}, \citenamefont {Mandrus},
  \citenamefont {Cheng}, \citenamefont {Jia}, \citenamefont {Wen},
  \citenamefont {Lee},\ and\ \citenamefont {Eom}}]{Putti10}%
  \BibitemOpen
  \bibfield  {author} {\bibinfo {author} {\bibfnamefont {M.}~\bibnamefont
  {Putti}}, \bibinfo {author} {\bibfnamefont {I.}~\bibnamefont {Pallecchi}},
  \bibinfo {author} {\bibfnamefont {E.}~\bibnamefont {Bellingeri}}, \bibinfo
  {author} {\bibfnamefont {M.~R.}\ \bibnamefont {Cimberle}}, \bibinfo {author}
  {\bibfnamefont {M.}~\bibnamefont {Tropeano}}, \bibinfo {author}
  {\bibfnamefont {C.}~\bibnamefont {Ferdeghini}}, \bibinfo {author}
  {\bibfnamefont {A.}~\bibnamefont {Palenzona}}, \bibinfo {author}
  {\bibfnamefont {C.}~\bibnamefont {Tarantini}}, \bibinfo {author}
  {\bibfnamefont {A.}~\bibnamefont {Yamamoto}}, \bibinfo {author}
  {\bibfnamefont {J.}~\bibnamefont {Jiang}}, \bibinfo {author} {\bibfnamefont
  {J.}~\bibnamefont {Jaroszynski}}, \bibinfo {author} {\bibfnamefont
  {F.}~\bibnamefont {Kametani}}, \bibinfo {author} {\bibfnamefont
  {D.}~\bibnamefont {Abraimov}}, \bibinfo {author} {\bibfnamefont
  {A.}~\bibnamefont {Polyanskii}}, \bibinfo {author} {\bibfnamefont {J.~D.}\
  \bibnamefont {Weiss}}, \bibinfo {author} {\bibfnamefont {E.~E.}\ \bibnamefont
  {Hellstrom}}, \bibinfo {author} {\bibfnamefont {A.}~\bibnamefont {Gurevich}},
  \bibinfo {author} {\bibfnamefont {D.~C.}\ \bibnamefont {Larbalestier}},
  \bibinfo {author} {\bibfnamefont {R.}~\bibnamefont {Jin}}, \bibinfo {author}
  {\bibfnamefont {B.~C.}\ \bibnamefont {Sales}}, \bibinfo {author}
  {\bibfnamefont {A.~S.}\ \bibnamefont {Sefat}}, \bibinfo {author}
  {\bibfnamefont {M.~A.}\ \bibnamefont {McGuire}}, \bibinfo {author}
  {\bibfnamefont {D.}~\bibnamefont {Mandrus}}, \bibinfo {author} {\bibfnamefont
  {P.}~\bibnamefont {Cheng}}, \bibinfo {author} {\bibfnamefont
  {Y.}~\bibnamefont {Jia}}, \bibinfo {author} {\bibfnamefont {H.~H.}\
  \bibnamefont {Wen}}, \bibinfo {author} {\bibfnamefont {S.}~\bibnamefont
  {Lee}}, \ and\ \bibinfo {author} {\bibfnamefont {C.~B.}\ \bibnamefont
  {Eom}},\ }\href {http://stacks.iop.org/0953-2048/23/i=3/a=034003} {\bibfield
  {journal} {\bibinfo  {journal} {Supercond. Sci. Technol.}\ }\textbf {\bibinfo
  {volume} {23}},\ \bibinfo {pages} {034003} (\bibinfo {year}
  {2010})}\BibitemShut {NoStop}%
\bibitem [{\citenamefont {Gurevich}(2014)}]{Gurevich14}%
  \BibitemOpen
  \bibfield  {author} {\bibinfo {author} {\bibfnamefont {A.}~\bibnamefont
  {Gurevich}},\ }\href {\doibase 10.1146/annurev-conmatphys-031113-133822}
  {\bibfield  {journal} {\bibinfo  {journal} {Annu. Rev. Cond. Matt. Phys.}\
  }\textbf {\bibinfo {volume} {5}},\ \bibinfo {pages} {35} (\bibinfo {year}
  {2014})}\BibitemShut {NoStop}%
\bibitem [{\citenamefont {Mikitik}\ and\ \citenamefont
  {Brandt}(2003)}]{Mikitik03}%
  \BibitemOpen
  \bibfield  {author} {\bibinfo {author} {\bibfnamefont {G.~P.}\ \bibnamefont
  {Mikitik}}\ and\ \bibinfo {author} {\bibfnamefont {E.~H.}\ \bibnamefont
  {Brandt}},\ }\href {\doibase 10.1103/PhysRevB.68.054509} {\bibfield
  {journal} {\bibinfo  {journal} {Phys. Rev. B}\ }\textbf {\bibinfo {volume}
  {68}},\ \bibinfo {pages} {054509} (\bibinfo {year} {2003})}\BibitemShut
  {NoStop}%
\bibitem [{\citenamefont {Kuo}\ \emph {et~al.}(2016)\citenamefont {Kuo},
  \citenamefont {Chu}, \citenamefont {Palmstrom}, \citenamefont {Kivelson},\
  and\ \citenamefont {Fisher}}]{Kuo16}%
  \BibitemOpen
  \bibfield  {author} {\bibinfo {author} {\bibfnamefont {H.-H.}\ \bibnamefont
  {Kuo}}, \bibinfo {author} {\bibfnamefont {J.-H.}\ \bibnamefont {Chu}},
  \bibinfo {author} {\bibfnamefont {J.~C.}\ \bibnamefont {Palmstrom}}, \bibinfo
  {author} {\bibfnamefont {S.~A.}\ \bibnamefont {Kivelson}}, \ and\ \bibinfo
  {author} {\bibfnamefont {I.~R.}\ \bibnamefont {Fisher}},\ }\href {\doibase
  10.1126/science.aab0103} {\bibfield  {journal} {\bibinfo  {journal}
  {Science}\ }\textbf {\bibinfo {volume} {352}},\ \bibinfo {pages} {958}
  (\bibinfo {year} {2016})}\BibitemShut {NoStop}%
\bibitem [{\citenamefont {Bud'ko}\ \emph {et~al.}(2009)\citenamefont {Bud'ko},
  \citenamefont {Ni},\ and\ \citenamefont {Canfield}}]{Sergey09}%
  \BibitemOpen
  \bibfield  {author} {\bibinfo {author} {\bibfnamefont {S.~L.}\ \bibnamefont
  {Bud'ko}}, \bibinfo {author} {\bibfnamefont {N.}~\bibnamefont {Ni}}, \ and\
  \bibinfo {author} {\bibfnamefont {P.~C.}\ \bibnamefont {Canfield}},\ }\href
  {\doibase 10.1103/PhysRevB.79.220516} {\bibfield  {journal} {\bibinfo
  {journal} {Phys. Rev. B}\ }\textbf {\bibinfo {volume} {79}},\ \bibinfo
  {pages} {220516} (\bibinfo {year} {2009})}\BibitemShut {NoStop}%
\end{thebibliography}

%

\end{document}